\documentclass[utf8]{FrontiersinHarvard}
\usepackage{url,hyperref,lineno,microtype,subcaption}
\usepackage[onehalfspacing]{setspace}

\usepackage{graphicx}

\usepackage{amsmath}
\usepackage{amsfonts}

\usepackage{mathtools}
\usepackage{pifont} 
\usepackage{wrapfig}
\usepackage{mathrsfs}
\usepackage{xcolor}
\usepackage{gensymb}
\usepackage[normalem]{ulem} %for \sout
\usepackage{bbm} % for all-one-vector

\newcommand{\dt}[1]{\textcolor{black}{#1}}
\newcommand{\dtr}[1]{\textcolor{black}{#1}}
\newcommand{\am}[1]{\textcolor{black}{#1}}
\newcommand{\amr}[1]{\textcolor{black}{#1}}
\newcommand{\mo}[1]{\textcolor{black}{#1}}
\newcommand{\dg}[1]{\textcolor{black}{#1}}
\newcommand{\kh}[1]{\textcolor{black}{#1}}
\newcommand{\bequ}{\begin{equation}}
\newcommand{\eequ}{\end{equation}}
\newcommand{\bequd}{\begin{eqnarray*}}
\newcommand{\eequd}{\end{eqnarray*}}

\newcommand{\td}{\textendash}

\def\P{\mathcal{P}}

\def\K{\mathcal{K}}

\def\R{\mathbb{R}}

\def\Bila{\mathbf{B}}

\def\keyFont{\fontsize{8}{11}\helveticabold }
\def\firstAuthorLast{Macarie {et~al.}} %use et al only if is more than 1 author
\def\Authors{Andrei Ciprian Macarie\,$^{1,3}$, Szabolcs Suveges\,$^{3}$, Mohamed Okasha\,$^{2}$, Kismet Hossain\textendash Ibrahim\,$^{2}$, J. Douglas Steele\,$^{3}$ and Dumitru Trucu \,$^{1,*}$.}
\def\Address{$^{1}$Division of Mathematics, University of Dundee, Dundee, United Kingdom \\
$^{2}$Department of Neurosurgery, Ninewells Hospital and School of Medicine, NHS Tayside, Dundee, United Kingdom\\
$^{3}$Division of Imaging Science and Technology, School of Medicine, University of Dundee, Dundee, United Kingdom}
\def\corrAuthor{Dumitru Trucu, Division of Mathematics, University of Dundee, Dundee, DD1 4HN, United Kingdom}

\def\corrEmail{trucu@maths.dundee.ac.uk}

\begin{document}
\onecolumn
\firstpage{1}

\title[GBM in oedema]{Post\textendash operative glioblastoma cancer cell distribution in the peritumoural oedema} 

\author[\firstAuthorLast ]{\Authors} %This field will be automatically populated
\address{} %This field will be automatically populated
\correspondance{} %This field will be automatically populated

\extraAuth{}% If there are more than 1 corresponding author, comment this line and uncomment the next one.
%\extraAuth{corresponding Author2 \\ Laboratory X2, Institute X2, Department X2, Organization X2, Street X2, City X2 , State XX2 (only USA, Canada and Australia), Zip Code2, X2 Country X2, email2@uni2.edu}

\maketitle

\begin{abstract}

\am{Glioblastoma multiforme (GBM), the most aggressive primary brain tumour, 
exhibits low survival rates due to its rapid growth, \dg{infiltrates} surrounding 
brain tissue, and \dg{is highly resistant} to treatment. One major challenge is oedema infiltration, 
a fluid build-up that provides a path for cancer cells to invade other areas. 
MRI resolution is insufficient to detect these infiltrating cells, leading to 
relapses despite chemotherapy and radiotherapy. In this work, we propose a new 
multiscale mathematical modelling \dg{method}, to explore the oedema
infiltration and predict tumour relapses. To address tumour relapses, we \dg{investigated} several
possible scenarios for the distribution of remaining GBM cells within the oedema after
surgery. Furthermore, in this computational modelling \dg{investigation on} tumour 
relapse scenarios \dg{were} investigated assuming the presence of clinically relevant 
chemo-radio therapy\kh{,
numerical} results suggest that a higher concentration of
GBM cells near the surgical cavity edge led to limited spread and slower progression of
tumour relapse. Finally, we explore mathematical and computational avenues for reconstructing relevant
shapes for the initial distributions of GBM cells within the oedema from available MRI scans.
The results obtained show good overlap between our simulation and the 
patient's \kh{serial MRI scans} taken 881 days into the treatment. While still under analytical investigation, this \dg{work} paves the way
for robust reconstruction of tumour relapses from available clinical data.}

\tiny
 \keyFont{ \section{Keywords:} multiscale modelling; cancer invasion; glioblastoma; chemotherapy; radiotherapy; surgery; 3D computational modelling; MRI scans} 
%All article types: you may provide up to 8 keywords; at least 5 are mandatory.
\end{abstract}

\section{Introduction}

%For Original Research Articles \citep{conference}, Clinical Trial Articles \citep{article}, and Technology Reports \citep{patent}, the introduction should be succinct, with no subheadings \citep{book}. For Case Reports the Introduction should include symptoms at presentation \citep{chapter}, physical exams and lab results \citep{dataset}.

Glioblastoma multiforme (GBM) is a devastating and highly invasive brain tumour that presents a significant treatment challenge. Despite the best efforts of medical professionals, the 5\textendash year survival rate for patients with GBM is only 7.2\% \citep{burri_2018,Wu_2021}. \dg{To} improve treatment outcomes, researchers have been exploring new approaches to tackling this aggressive disease. One promising avenue of investigation is the use of mathematical models to simulate tumour evolution and explore potential new treatment strategies \citep{trucu_2013_multiscale,michor_2015, malinzi_2017_modelling,shuttleworth_2019_multiscale,suveges_2021_mathematical}.

GBM is typically treated with surgical resection if possible, followed by chemoradiotherapy. The Stupp protocol \kh{is the standard of care} treatment regimen that involves a total of 60 grays \am{(abbreviated as \textit{Gy}, and representing the unit of measurement for absorbed radiation)} of radiotherapy delivered in daily doses of 2 Gy over 6 weeks, along with the chemotherapy drug Temozolomide (TMZ). During radiotherapy, patients take 75 mg of TMZ per square meter of body surface area every day for 7 days a week. After radiotherapy is completed, TMZ (adjuvant) is given in 6 cycles of \am{150\textendash 200} mg per square meter for 5 days \kh{every 28 days \citep{Stupp_2005}.} 

	Treating GBM is a formidable challenge due to several factors. Even after maximal surgical resection and adherence to the Stupp protocol, approximately 90\% of patients experience local recurrences \citep{Heterogeneity_pte, benefit_recurrent, bayesian_meta, pattern_failure}. Another significant challenge is the high infiltration and heterogeneity of GBM, which makes it difficult to identify tumour margins accurately. GBM grows with microscopic \am{finger\textendash like} projections that extend beyond what MRI scans (the gold standard for brain tumour imaging) can detect \citep{Wu_2021}. Furthermore, GBM cells invade the brain through the peritumoural oedema (PTE), a condition in which fluid accumulates in the extracellular spaces of brain tissue surrounding the tumour. PTE is formed by tumour cells, reactive astrocytes, and inflammatory cells. The infiltrating GBM cells in the PTE are phenotypically distinct from those isolated from the corresponding mass. Residual GBM cells located at the resection margin are known to proliferate more quickly and be more invasive than GBM cells found in the tumour center \kh{\citep{Heterogeneity_pte, Qin_2021}}. Therefore, it is crucial to examine the PTE, as it could lead to recurrences since this area is not \kh{always treated \citep{Niyazi_2023}}.

	The limited effectiveness of traditional GBM treatments underscores the need for innovative approaches \mo{\citep{michor_2015, Yalamarty_2023}}. In recent years, mathematical models have emerged as a promising tool for gaining insights into GBM tumour growth and progression \citep{hatzikirou_2005_mathematical,swanson_2008,Rockne_2010,Radio_Bayesian_GBM,suveges_2021_mathematical,Plaszczynski_2023}. By incorporating clinical data and biological parameters, mathematical models can provide a more comprehensive understanding of tumour behaviour than traditional experimental techniques alone \citep{Le_2017,Review_Solid_Tumours}. However, most of these studies are limited to simulating tumour growth in two dimensions and on one spatio\textendash temporal scale \citep{suveges_2021_mathematical}. Nevertheless, significant progress has been made in developing multiscale moving boundary modelling and computational frameworks for tumour growth \citep{trucu_2013_multiscale,shuttleworth_2019_multiscale,suveges_2021_mathematical}. As detailed below, the combination of these modelling approaches paves the way for the work discussed \dg{here}.

	In this work we aim to explore the distribution of GBM cells within the oedema. The underlying motivation for this is \dt{the understanding of the relationship between the spatial distribution of cancer cells within oedema that remain post\textendash surgery and the likelihood of post\textendash surgical tumour recurrence. This will combine novel mathematical multiscale moving boundary modelling with \dg{NHS} clinical data assimilation using MRI scans from a single patient with diagnosed GBM.} We explore two scenarios: the first utilizes a standard mollifier to describe cell distribution inside the oedema, while the second uses a Gaussian distribution.

\am{This paper presents a multiscale moving boundary model for simulating GBM evolution, incorporating treatment effects and clinical data.
After introducing our multiscale modelling for GBM progression, we formulate our tumour relapse hypothesis and outline \amr{the mathematical and computational strategy for clinical data inversion (\emph{i.e.,} assimilate MRI images within our modelling to enable tumour recurrence predictions)}. \amr{Details about prospectively collected MRI scans (from GBM patients at Ninewells Hospital) alongside their pre\textendash processing pipeline are also included}. The manual tumour segmentation was carried out under the supervision of consultant neurosurgeons, Mr. Kismet Ibrahim \mo{(referred here as KHI)} and Mr. Mohamed Okasha \mo{(referred here as MO)}. \amr{Finally, we describe the multiscale numerical scheme involved in approximating the mathematical model computationally, and present the simulation results as well as discuss future research avenues.}}

\section{Materials and Methods}
\am{This section details the mathematical model that we developed to simulate the evolution of \dg{GBM} within a three\textendash dimensional fibrous brain environment. Our framework \dg{expands} the work of \citet{suveges_2021_mathematical} by incorporating the effects of various treatment modalities such as surgery, chemotherapy, and radiotherapy. Furthermore, we postulate our hypothesis and formulate a minimisation problem.   Finally, we leverage clinical data from T1, T1+C, T2, and DTI scans to account for factors like brain structure, tumour location and extent, and oedema.}
%\begin{figure}[ht!]
%\centering
%\begin{subfigure}[t]{0.49\textwidth}
%        \raisebox{-\height}{\includegraphics[width=0.49\textwidth]{T1CGBM1}}
%        \raisebox{-\height}{\includegraphics[width=0.49\textwidth]{T1CGBM1VOI}}
%        \vspace{.6ex}
%        \raisebox{-\height}{\includegraphics[width=0.49\textwidth]{T2P1}}
%        \raisebox{-\height}{\includegraphics[width=0.49\textwidth]{EdemaP1}}
%
%\end{subfigure}
%\caption{a) T1+C, b) GBM VOI, c) T2 and d) oedema VOI scans.}
%\label{GBMinScans}
%\end{figure}

\subsection{\am{Mathematical Multiscale Model for GBM Progression}} 

\subsubsection{Macro\textendash scale dynamics}

Following the work from \citet{trucu_2013_multiscale,shuttleworth_2020,suveges_2021_mathematical}, we denote by $\Omega(t)$ the expanding 3\textendash dimensional (3D) tumour region that progresses over the time interval $[0, T]$ within a maximal tissue cube $Y \subset \mathbb{R}^{3}$. %Create a schematic like Figure 1 in Szabolcs' paper. 
At any macro\textendash scale spatio\textendash temporal point $(x, t)\in Y \times [0, T]$, we consider a cancer cell population, denoted by $c(x, t)$, which interacts with a two\textendash phase heterogeneous ECM (consisting of: a  non\textendash fibre $l(x, t)$ and fibre $F(x, t)$ ECM phases \citep{suveges_2021_mathematical}), while consuming the available nutrients, denoted by $\sigma(x,t)$, \dtr{which are present in the environment.} The fibre ECM density, $F(x, t)$, accounts for all fibrous proteins such as collagen and fibronectin. On the other hand, the non\textendash fibre ECM density, $l(x, t)$, \dg{comprises} of non\textendash fibrous proteins (for example, amyloid fibrils), extracellular $Ca^{2+}$ ions, enzymes and polysaccharides \citep{suveges_2021_mathematical}. Following the methods introduced in \citet{suveges_2021_mathematical}, we also incorporate the structure of the brain by extracting data from the modified DTI scan, T1 and T2 brain scans. Finally, we denote by $\mathbf{u}(x,t)$ the global tumour vector which embodies the \dg{cancer cell population and} the fibre and non\textendash fibre ECM components, given by

\begin{equation*}
\mathbf{u}(x,t) := (c(x, t), l(x, t), F(x, t))^T.
\end{equation*}

Therefore, the total space occupied by the macroscopic tissue and tumour volume is denoted by $\rho(\mathbf{u})$ and is defined as

\begin{equation*}
\rho(\mathbf{u}) = \rho(c(x, t), l(x, t), F(x, t)) := c(x, t) + l(x, t) + F(x, t),
\end{equation*}

for all $(x,t)\in \Omega(t)\times [0,T]$. 

\paragraph{Nutrients:} \dtr{As in this study we} focus on avascular tumours, \dtr{the uptake of nutrients that are available in the outside tissue and are absorbed through the outer tumour boundary plays an important role in the overall tumour development. This nutrients absorption is assumed here to occur at the constant rate $d_\sigma>0$ and is enabled in the model through the presence of nutrient Dirichlet boundary condition at the evolving tumour boundary $\partial \Omega(t)$. Furthermore, the spatio-temporal nutrient transport is assumed to be in diffusion equilibrium, with an autonomous transport diffusion coefficient $\mathcal{D}_\sigma=D_\sigma/(c+F+p_0)$ that takes account of both the presence of the cancer and ECM fibres distributions as well as the baseline permeability $p_0>0$ (which is here assumed to be a media constant), while  $D_\sigma>0$ is a constant standing for the maximal diffusive nutrients transport possible in the tissue. Thus, the nutrients dynamics is mathematically given by: } 
\begin{equation}
\label{Nutrient_Eq}
\begin{array}{rll}
0 =& \nabla \cdot (\mathcal{D}_\sigma \nabla)\sigma - d_\sigma c \sigma, &\textrm{ on } \Omega(t), \forall t \in [0,T],\\
\sigma(x,t)=&\sigma_{nor}, &\forall x \in \partial \Omega_{0}(t), \forall t \in [0,T],
\end{array}
\end{equation}
\dtr{where $\sigma_{nor}$ is the normal level of nutrients in the outside tissue and is considered to be constant, while \amr{$\partial\Omega_{0}(t)$} represents the outside tumour boundary as defined in Appendix \ref{outTumourBry}.} Similar to \citet{Szabolcs_2022_Nutrients}, \dtr{certain tumour regions become necrotic as soon as the nutrients level $\sigma$ drop below a critical necrotic threshold denoted $\sigma_n>0$, while $\sigma_{p}>0$ represents a nutrient for optimal cancer proliferation regime.} Hence, we have the following relationship between these three values: $\sigma_{nor}>\sigma_p>\sigma_n$.

\dtr{Further, considering here a simpler context than the one in  \citet{Szabolcs_2022_Nutrients} by focussing only on two nutrient effects (namely, on cell proliferation and cell death rates), we assume that: (1)  very low nutrient levels impede cell proliferation (having no proliferation at all in the necrotic regions); and (2) extremely high nutrient levels cannot increase cell proliferation rate by more than} a certain maximal proliferation rate $\Psi_{p,max} > 0$ which corresponds to nutrient levels $\sigma \geq \sigma_p$. Thus, \dtr{mathematically, these two assumptions are accounted for in the modelling via the following nutrient\textendash dependent proliferation function:}
\begin{equation}
\label{Nutrient_proliferation}
\Psi_p(\sigma) := \begin{cases}
0, & \text{if } \sigma \leq \sigma_n, \\
\Psi_{p,max}, & \text{if } \sigma \geq \sigma_p, \\
\Phi(\sigma, \Psi_{p,max},0,\sigma_p-\sigma_n), & \text{otherwise},
\end{cases}
\end{equation}
where $\Phi(\sigma,.,.,.)$ describes the smooth transition between the two extrema and is defined to be:
\begin{equation}
\label{Transition_Eq}
\Phi(\sigma, \Phi_{max},\Phi_{min},\Phi_L) := \frac{\Phi_{max}-\Phi_{min}}{2} \Bigl [cos\Bigl(\frac{\pi (\sigma-\sigma_n-\Phi_L)}{\sigma_p-\sigma_n}\Bigr)+1\Bigr] + \Phi_{min},
\end{equation}
where $\Phi_L$ controls the phase shift of the cosine function.

\dtr{Finally, the effect that the nutrients absence/presence have on cancer cell death is characterised via a function $\Psi_{d}(\sigma)$ that is of similar type as the one given in \amr{Equation} \eqref{Nutrient_proliferation}. Specifically, here we consider a maximal death rate $\Psi_{d,max}>0$ in necrotic regions, while} we assume no death for cancerous cells when the level of nutrients is $\sigma \geq \sigma_p$. \dtr{Thus, using again the transition function from Equation \eqref{Transition_Eq}, the effect over the death rate of cancer cells is mathematically expressed as:}
\begin{equation}
\label{Nutrient_death}
\Psi_{d}(\sigma) := \begin{cases}
\Psi_{d,max}, & \text{if } \sigma \leq \sigma_n, \\
0, & \text{if } \sigma \geq \sigma_p, \\
\Phi(\sigma, \Psi_{d,max},0,0), & \text{otherwise}.
\end{cases}
\end{equation}

\paragraph{Cancer cell dynamics: $c(x,t)$.}

The \dtr{spatio\td temporal dynamics of the cancer cell population considered in this work accounts for available movement characteristics enabled by} T1 and DTI scans \citep{ref53}, \dtr{based on which the fully anisotropic diffusion tensor, denoted by $\mathbb{D}_T$ \citep{suveges_2021_mathematical,ref37,ref65,ref66}. In addition to that, the cell population movement is further biased by adhesion processes, which are mathematically captured through a term denoted by $\mathcal{A}(x,t,\mathbf{u},\theta_f)$ that will be detailed below.} Furthermore, we assume a logistic type proliferation law of the form:
\begin{equation}
P(\mathbf{u}):=\mu \Psi_p(\sigma) c(1-\rho(\mathbf{u}))^{+},
\end{equation}
where $\mu>0$ is the proliferation rate regulated by the available nutrients, represented here by the nutrient proliferation function $\Psi_p(\sigma)$ given by Equation \eqref{Nutrient_proliferation}. Additionally, the term $(1-\rho(\mathbf{u}))^+$ guarantees that we \amr{do not experience cell population overcrowding within the available space.}
 
	\dtr{Further, while it is well known that one of the hallmarks of cancer is resisting death \citep{Hallmarks_Cancer_2022}, nevertheless, due to the \kh{abnormal} peritumoural vasculature and the degradation of the ECM, nutrient delivery is reduced inside the tumour,  ultimately leading to necrosis \citep{Szabolcs_2022_Nutrients}. Therefore, we assume a death rate $d>0$ that is regulated by the cancer cell death function $\Psi_{d}(\sigma)$ given by Equation \eqref{Nutrient_death}. Thus, mathematically the cancer cell death is captured here by the term}:
\begin{equation}
Q(\mathbf{u}):=d \Psi_{d}(\sigma)c.
\end{equation}	
Finally, the population of cancer cells is being reduced \dtr{further} by the effects of chemotherapy and radiotherapy\kh{, which are cross\textendash referenced with the patient's post treatment MRI scans}. \dtr{Hence, the spatio\td temporal cancer population dynamics is given mathematically by the following partial differential equation:} 
\dt{\bequ\label{cancerDynamics}
\begin{array}{cll}
 \frac{\partial c}{\partial t} &=& \underbrace{\nabla\nabla:[\mathbb{D}_{T}(x)c]}_{\substack{\text{Diffusion}}} - \underbrace{\nabla[c \mathcal{A}(x,t,\mathbf{u},\theta_f)]}_{\substack{\text{Adhesion interactions}}}+ P(\mathbf{u})-Q(\mathbf{u})\\[0.8cm]
 &&-Radiotherapy(c,t)-Chemotherapy(c,t).
 \end{array}
\eequ}\amr{The first term in Equation \eqref{cancerDynamics}, $\nabla\nabla:[\mathbb{D}_{T}(x)c]$, denotes the full second order anisotropic tumour diffusion, with the 3D  diffusion tensor $\mathbb{D}_T$ being constructed from DTI scans of the brain \citep{engwer_2014_glioma,suveges_2021_mathematical} and ultimately given by}:
\begin{multline}
 \mathbb{D}_{T}(x):=D_{c}D_{WG}(x)\Bigl[\left(r+(1-r)\left(\frac{\coth k (x)}{k(x)}-\frac{1}{k(x)^2}\right)\right)I_{3}\\+(1-r)\left(1-\frac{3\coth k(x)}{k(x)}+\frac{3}{k(x)^2}\right)\phi_{1}(x)\phi_{1}^T(x)\Bigr].
\end{multline}
\dtr{Here, $D_c>0$ is the diffusion coefficient, while $D_{WG}(\cdot)$ acts as a regulator term, addressing the well known fact that malignant glioma cells have higher motility in white matter than in grey matter \amr{\citep{Chicoine_1995, Silbergeld_1997,Swanson_2000, Brooks_2021}}, and is defined as:
\begin{equation}
D_{WG}(x)=\bigl((D_{_{G}} g(x)+w(x))\ast\psi_\rho \bigr)(x),
\end{equation}
The ratio between the motility regimes in grey and white matter is given here by $D_{_{G}} \in[0,1]$, $g(x)$ and $w(x)$ are the grey and white matter densities, respectively, which are obtained from the T1 scan \citep{suveges_2021_mathematical}. Further, $\psi_\rho(x):=\psi_{_{3}}(x/\rho)/\rho^3$ is the mollifier induced by the standard mollifier $\psi_{_{3}}$ defined in Appendix \ref{mollifierAppendix}, and $\ast$ denotes the convolution operator.
Furthermore, $r\in [0,1]$ is the \dtr{extent} of isotropic diffusion, $I_3$ is the $3\times 3$ identity matrix. Moreover, $\lambda_1(x)\geq \cdots \geq \lambda_N(x)$ denote the eigenvalues, while $\phi_1(x), \cdots,\phi_N(x)$ represent the corresponding eigenvectors. Finally,} $k(x)$ is given by
\bequd
k(x):=\K_{FA} FA(x),
\eequd
with $\mathcal{K}_{FA}\geq 0$ measuring the sensitivity of the cells to the direction of the environment, \dtr{while $FA(x)$ stands for} the \emph{fractional anisotropy index} \citep{engwer_2014_glioma,suveges_2021_mathematical} \dtr{and is defined as}
\bequd
FA(x):=\sqrt{\frac{(\lambda_1(x)-\lambda_2(x))^2+(\lambda_2(x)-\lambda_3(x))^2+(\lambda_1(x)-\lambda_3(x))^2}{2(\lambda_1^2(x)+\lambda_2^2(x)+\lambda_3^2(x))}}.
\eequd
The second term \dt{in Equation \eqref{cancerDynamics}, namely} $\nabla[c \mathcal{A}(x,t,\mathbf{u},\theta_f)]$, describes adhesion processes that bias the movement of the cell \dt{population due to the adhesion bonds that the migratory cells establish with both the surrounding cell and the ECM components}. Introduced in \citet{shuttleworth_2019_multiscale} and expanded later in \citet{suveges_2021_mathematical}, the \am{non\textendash local} flux term considers the interactions of cancer cells within a sensing region $\mathbf{B}(0,R)$, with radius $R>0$, described by:
\begin{multline} \label{non_flux term}
\mathcal{A}(x,t,\mathbf{u},\theta_f):=\frac{1}{R} \int_{\mathbf{B}(0,R)} \mathcal{K}(y)\Bigl[n(y)(\mathbf{S}_{cc}c(x+y,t)+\mathbf{S}_{cl}l(x+y,t)) \\ + \hat{n}(y,\theta_f(x+y,t))\mathbf{S}_{cF}(x+y,t)\Bigr][1-\rho(\mathbf{u})]^{+}dy,
\end{multline}
where $\mathbf{S}_{cc}, \mathbf{S}_{cl}, \mathbf{S}_{cF}>0$ are the cell\textendash cell, cell\textendash non\textendash fibrous ECM and cell\textendash fibrous ECM adhesion strength coefficients, respectively. $\mathbf{S}_{cc}$ is positively correlated to the levels of extracellular $Ca^{2+}$ ions. Hence, we describe the cell\textendash cell bonds as:
\bequd
\mathbf{S}_{cc}:=\mathbf{S}_{\min}+(\mathbf{S}_{\max}-\mathbf{S}_{\min})\exp \Bigl[1-\frac{1}{1-(1-l(x,t))^{2}}\Bigr],
\eequd
with $\mathbf{S}_{\min}>0$ and $\mathbf{S}_{\max}>0$ are the minimum and maximum levels of $Ca^{2+}$ ions \citep{suveges_2021_mathematical,Szabolcs_2022_Nutrients}.
Furthermore, the gradual weakening of these bonds are represented by using a radially symmetric kernel $\mathcal{K}(\cdot)$ given by:
\bequd
\mathcal{K}(y)=\psi_{_{1}}\Big(\frac{y}{R}\Big), \qquad\forall y \in \mathbf{B}(0,R),
\eequd
where $\psi_{_{1}}(\cdot)$ is the standard mollifier defined in Appendix \ref{mollifierAppendix}.
Moreover, in Equation \eqref{non_flux term}, $n(\cdot)$ and $\hat{n}(\cdot,\cdot)$ are the unit radial vector and unit radial vector biased by the oriented ECM fibres \citep{suveges_2021_mathematical}, described mathematically as
\begin{align*}
n(y)&:=\begin{cases} \frac{y}{\|y\|_{_{2}}} & \mbox{if } y\in \mathbf{B}(0,R) \setminus \{0\}, \\ 0 & \mbox{if } y=0,\end{cases}
\\
\hat{n}(y,\theta_f(x+y,t))&:=\begin{cases} \frac{y+\theta_f(x+y,t)}{\|y+\theta_f(x+y,t)\|_{_{2}}} & \mbox{if } y\in \mathbf{B}(0,R) \setminus \{0\}, \\ 0 & \mbox{if } y=0.\end{cases}
\end{align*}
Finally, to prevent overcrowded regions contributing to cell migration, we have a limiting term $[1-\rho(\mathbf{u})]^{+}:=\max(0,1-\rho(\mathbf{u}))$ \citep{suveges_2021_mathematical}.		

	\amr{The governing equation also accounts for the effects of radiotherapy and chemotherapy}. Radiotherapy \dt{is administered in multiple sessions scheduled according to \dg{five days a week sequence (Monday to Friday) in equal amounts of doses} \amr{that is captured here mathematically via} a subsequence of days $\{j_{_{m}}\}_{_{k=1\dots N_{radio}}}\subset \{1, \dots, N_{final}\}$} (where $\{1, \dots, N_{final}\}$ represents the entire period of treatment).
	 \dt{The intensity of each radiotherapy fraction follows the} \am{linear\textendash quadratic} model introduced in \citet{Irina_2021} and is delivered here according to an appropriate per-day radiotherapy distribution function $\bar r:\{1, \dots, N_{_{radio}}\}\to (0,\infty)$, given by \am{$\bar r(j_{_{m}})=\alpha D(j_{_{m}}) + \zeta D(j_{_{m}})^2$, where $\alpha>0$ and $\zeta>0$ are linear and quadratic coefficients of cell damage,} and $D(\cdot):\{1, \dots, N_{_{radio}}\}\to (0,\infty)$ is the per-day radiation dose level distribution (\emph{i.e.,} indicating the dose administered in each scheduled day). \dt{Finally, \dg{we account} here also \dg{for} the time\textendash overlapping} effect of radiotherapy treatment over each time interval $(T_{_{i_{_{k}}}}-l,T_{_{i_{_{k}}}}+d)$ \dt{via the asymmetric mollifier-type function  $\psi^{radio}_{j_{_{m}}}(t)$ given in Appendix~\ref{mollifierAppendix}, Equation \eqref{schedulingFunc}, $\forall m\in \{1, \dots, N_{_{radio}}\}$, we have that mathematically the radiotherapy treatment delivery and its effect on the tumour \dg{is given} by
	\bequ	Radiotherapy(c(x,t),t):=\displaystyle\sum_{m=1}^{N_{radio}} \bar r(j_{_{m}})\psi^{radio}_{j_{_{m}}}(t) c(x,t).
	\eequ}Chemotherapy \amr{is incorporated based on the} \am{Norton\textendash Simon} hypothesis \citep{Irina_2021}, which \am{suggests that tumours are more susceptible to treatment when they have grown for a shorter period of time.} \dt{Following a chemotherapy scheduling given by a selected subsequence of days \am{$\{i_{_{k}}\}_{_{k=1\dots N_{chemo}}}\subset \{1, \dots, N_{final}\}$},} \dt{we deliver $N_{_{chemo}}$ doses of chemotherapeutic drug,  according to the corresponding per\textendash day chemo agent distribution function $\rho_g:\{1,\dots,N_{_{chemo}}\}\rightarrow \{1, 1.1, 1.5, 2, 2.4, 2.5, 2.8\}\times chemo_{_{dose}}$, with $chemo_{_{dose}}>0$ being the initial chemo dose. The time\textendash overlapping effect of the chemotherapy over the interval $(T_{_{i_{_{k}}}}-l,T_{_{i_{_{k}}}}+d)$ is accounted here via a function $\psi^{chemo}_{{i_{_{k}}}}(t)$, given in Appendix~\ref{mollifierAppendix}, Equation \eqref{schedulingFunc}, which is similar in shape to the one for radiotherapy. Further, to account for the fractional cell kill impaired by cytotoxic agent,} we adopt an Exponential Kill Model given by $b(1-e^{\beta W})$, where $b>0$ represents the relative maximum fractional cell kill, $W>0$ \dt{stands for} the drug concentration, and $\beta>0$ \dt{describes} tumour cells' sensitivity to the chemo drug. \dt{Moreover, the decrease in fractional cell kill as tumour cell population gets closer to its carrying capacity \amr{$K>0$ (representing the maximum cumulative distribution of cells and ECM supported by an infinitesimal volume of tissue)} is described here through} a Holling type II functional $\mu K/(K+sc)$, where $\mu>0$ is the growth rate, and $s>0$ controls the extent of the \am{Norton\textendash Simon} effect, \am{\emph{i.e.,} a larger $s$ leads to a steeper decline, effectively amplifying the Norton-Simon effect by significantly reducing cell kill effectiveness when the tumour is close to its capacity. Conversely, a smaller $s$ results in a more gradual decline, making the Norton-Simon effect less pronounced and allowing for potentially higher cell kill even at larger tumour sizes} \citep{Irina_2021}. \dt{Thus, chemotherapy delivery and its effect on the tumour is given mathematically by: 
\begin{equation}
Chemotherapy(c(x,t),t):=\mu b\frac{K}{(K+sc(x,t))}(1-e^{\beta W}) \sum_{k=1}^{N_{chemo}}\rho_{g}({i_{_{k}}}) \psi^{chemo}_{{i_{_{k}}}}(t)c(x,t)
\end{equation}}	 
Thus, the governing equation for cancer Dynamics finally becomes
\am{\bequ\label{c(x,t)}
\begin{array}{cll}
 \frac{\partial c}{\partial t} &=& \underbrace{\nabla\nabla:[\mathbb{D}_{T}(x)c]}_{\substack{\text{Diffusion}}} - \underbrace{\nabla[c \mathcal{A}(x,t,\mathbf{u},\theta_f)]}_{\substack{\text{Adhesion interactions}}}+ P(\mathbf{u})-Q(\mathbf{u})\\[0.8cm]
 &&- \underbrace{\displaystyle\sum_{m=1}^{N_{radio}} \bar r(j_{_{m}})\psi^{radio}_{{j_{_{m}}}}(t)c}_{\substack{\text{Radiotherapy}}}-\underbrace{\mu b \frac{K}{K+sc} (1-e^{\beta W}) \displaystyle\sum_{k=1}^{N_{chemo}} \rho_g({i_{_{k}}})\psi^{chemo}_{i_{_{k}}}(t)c}_{\substack{\text{Chemotherapy}}}.
 \end{array}
\eequ}
\paragraph{\am{Two\textendash Phase ECM macro\textendash scale} dynamics: $F(x,t)$ and $l(x,t)$.}
The \dtr{micro\textendash scale mass distribution of} fibre \dtr{ECM phase} determines \dtr{a} spatial orientation of ECM fibres at \am{micro\textendash scale} level \dtr{which represents their naturally emerging spatial bias for withstanding incoming cell forces \citep{shuttleworth_2019_multiscale}. With this} orientation, while deferring more consistent details for a later subsection, the ECM fibre phase is therefore represented as \dtr{a macroscopic} vector field $\theta_f(x,t)$ \dtr{ whose Euclidean norm stands for} the amount of fibres at a given \am{macro\textendash scale} point $(x,t)$, and so $F(x,t) :=\|\theta_f(x, t)\|_{_{2}}$ \citep{shuttleworth_2019_multiscale,suveges_2021_mathematical}.
	\dtr{Further,} to incorporate the impact of treatment on the \dtr{each of the two} ECM \dtr{phases}, we build on the dynamics of the fibre and \am{non\textendash fibre} ECM components introduced in \citet{suveges_2021_mathematical, Szabolcs_2022_Nutrients} \dtr{by considering also the decay effects that the chemo and radio therapies bring about, namely: 
\begin{equation} \label{F(x,t)}
\frac{\partial F}{\partial t} = -Fc(\beta_F +\beta_{FChemo}  + \beta_{FRadio}),
\end{equation}
\begin{equation} \label{l(x,t)}
\frac{\partial l}{\partial t} = -lc(\beta_l +\beta_{lChemo}+\beta_{lRadio}),
\end{equation}
where $\beta_{FChemo}, \beta_{FRadio}$ and $\beta_{lChemo},\beta_{lRadio}$ are the corresponding constant decay rates due to the chemo and radio therapies on the ECM fibres and non-fibres phases, respectively.}
\paragraph{Summary of the full macro\textendash scale model}In summary, the full model for the macro\textendash scale dynamics is:

\am{
\bequ
\left\{
\begin{array}{lll}
\frac{\partial c}{\partial t}&=&\nabla\nabla:[\mathbb{D}_{T}(x)c] - \nabla[c \mathcal{A}(x,t,\mathbf{u},\theta_f)]+ P(\mathbf{u})-Q(\mathbf{u})\\[0.5cm]
&&-\displaystyle\sum_{m=1}^{N_{radio}} \bar{r}(j_{_{m}})\psi^{radio}_{j_{_{m}}}(t) c-\mu b \frac{K}{K+sc} (1-e^{\beta W}) \displaystyle\sum_{k=1}^{N_{chemo}} \rho_g({i_{_{k}}})\psi^{chemo}_{i_{_{k}}}(t)c, \\ \\[0.1cm]
 \frac{\partial F}{\partial t}&=&-Fc(\beta_F +\beta_{FChemo} +\beta_{FRadio} ), \\ \\[0.1cm]
 \frac{\partial l}{\partial t}&=&-lc(\beta_l +\beta_{lChemo}+\beta_{lRadio}), \\ \\[0.1cm]
 0&=&\nabla \cdot (\mathcal{D}_\sigma \nabla)\sigma - d_\sigma c \sigma,
 \end{array}
 \right.
 \label{CompleteMacro}
\eequ}
in the presence of \am{zero\textendash flux} boundary conditions
for the cancer, fibre and non\textendash fibre ECM phases, as well as, Dirichlet boundary condition for the nutrients.

\subsubsection{\am{Micro\textendash scale} dynamics \dt{within the bulk and at the tumour boundary}}\label{Links}

%Cancer invasion is a complex process that involves multiple scales, from the individual cancer cell to the tumour as a whole. At the \am{micro\textendash scale}, cancer cells interact with the extracellular matrix (ECM) and with each other. Cancer cells can degrade the ECM using \am{matrix\textendash degrading} enzymes (MDEs), which allows them to invade surrounding tissues. At the \am{macro\textendash scale}, the tumour grows and spreads through the brain. The growth and spread of the tumour is influenced by the \am{micro\textendash scale} interactions between cancer cells and the ECM.

In this section, we focus on the \am{micro\textendash scale} processes that contribute to cancer invasion. We first discuss the rearrangement of ECM fibres by cancer cells. ECM fibres are important for providing structural support to tissues. Cancer cells can rearrange ECM fibres using matrix\textendash degrading enzymes (MDEs), such as matrix\textendash metalloproteinases, which allows them to create new pathways for invasion. We then discuss the \am{cell\textendash scale} proteolytic process at the edge of the tumour, \dtr{whereby} cancer cells secrete MDEs that degrade the ECM, allowing for further \dtr{tumour} invasion. \dtr{Finally}, we discuss the naturally arising double feedback loop that connects the \am{micro\textendash scale and macro\textendash scale}. In this loop, the \am{micro\textendash scale} interactions between cancer cells and the ECM influence the \am{macro\textendash scale} growth and spread of the tumour. The \am{macro\textendash scale} growth and spread of the tumour, in turn, influences the micro\textendash scale interactions between cancer cells and the ECM \citep{shuttleworth_2019_multiscale,suveges_2021_mathematical,Szabolcs_2022_Nutrients}.

\paragraph{Micro\textendash scale dynamics of ECM fibres and their macro\textendash scale implications.}

As described in  \citet{shuttleworth_2019_multiscale,suveges_2021_mathematical,Szabolcs_2022_Nutrients}, the macroscopic ECM fibres \dtr{alongside their ability to withstand incoming forces are represented through the vector field  $\theta_f (x,t)$ that at each spatio\textendash temporal node $(x,t)$ is non\textendash locally induced from their micro\textendash scale configuration as follows:} 
\begin{equation}
\theta_f (x,t):= \frac{1}{\lambda (\delta Y(x))} \int \limits_{\delta Y(x)} f(z,t)dz \cdot \frac{\theta_{f,\delta Y(x)}(x,t)}{\lVert \theta_{f,\delta Y(x)}(x,t) \rVert_2}.
\end{equation}
\dtr{Here, $f(z,t)$ is the micro\textendash scale mass density of micro\textendash fibres distributed on a micro\textendash domain $\delta Y(x):= x+\delta Y$ of appropriate micro\textendash scale size $\delta > 0$, while $\lambda(\cdot)$ is the usual Lebesque measure in $\mathbb{R}^3$.} Further, $\theta_{f,\delta Y(x)}(\cdot,\cdot)$ is the revolving barycentral orientation given by: 
\bequd
\theta_{f,\delta Y(x)} (x,t):= \frac{\int \limits_{\delta Y(x)} f(z,t)(z-x)dz}{\int \limits_{\delta Y(x)} f(z,t)dz}.
\eequd
\dtr{Thus, the global macro\textendash scale oriented ECM fibre $\theta_f (x,t)$ characteristics (including its Euclidean magnitude which represent the amount of fibres at $(x,t)$, namely $F(x,t):=\lVert \theta_{_{f}}(x,t) \rVert_{_{2}}$), arise and are fully determined from the micro\textendash scale distribution of ECM fibres, providing this way \textit{a fibres bottom\textendash up} \emph{micro\textendash to\textendash macro scales} link.}

\dtr{However,} there exists also \textit{a macro\textendash to\textendash micro scales fibres top\textendash bottom link, which} is triggered by the movement of cancer cells \dtr{through the ECM fibre distribution} that \dtr{cause the} rearrangement \dtr{of} the ECM micro\textendash fibres on each micro\textendash domain $\delta Y(x)$. \dtr{Specifically}, the fibre rearrangement process is \dtr{triggered by} the \dtr{macro\textendash scale} cancer cell spatial flux
\begin{equation}
\mathcal{F}(x,t):= \mathbb{D}_T(x)\nabla c + c\nabla \cdot \mathbb{D}_T(x)-c\mathcal{A}(x,t,\mathbf{u},\theta_f),
\label{Flux}
\end{equation}   
\dtr{which is balanced by the oriented macro\textendash scale ECM fibre $\theta_{_{f}}(x,t)$, resulting in a rearrangement flux
\begin{equation}
r(\delta Y(x),t):= w(x,t)\mathcal{F}(x,t)+(1-w(x,t))\theta_f(x,t).
\end{equation} 
with $w(x,t):=c(x,t)/(c(x,t)+F(x,t))$ being an appropriate mediating weight taking into account the amount of cells transported at $(x,t)$ relative to the overall amount of cells and fibres at $(x,t)$. This acts uniformly on the mass distribution of micro\textendash fibre on each micro\textendash domain $\delta Y(x)$, and induces a reallocation of the mass distribution of micro\textendash fibres within both $\delta Y(x)$ and its adjacent neighbouring micro\textendash domains, as described in  \citet{shuttleworth_2019_multiscale,suveges_2021_mathematical,Szabolcs_2022_Nutrients}.}

\paragraph{MDE\dt{s boundary} micro\textendash dynamics and its links \dt{to the macro\textendash dynamics}}

\dtr{Besides the bulk micro\textendash dynamics that involve the ECM fibres, another key micro\textendash dynamics for tumour invasion is the one involving the proteolytic activity that occurs on the invasive edge of the tumour, enabled by the MDEs (secreted by the cancer cells close to the tumour interface) and transported within the surrounding cell\textendash scale peritumoural tissue neighbourhood.} Consequently, \dtr{this MDE micro\textendash dynamics cause} degradation of the peritumoural ECM, thereby inducing alterations in the morphological contours of the tumour boundary \citep{suveges_2021_mathematical,Szabolcs_2022_Nutrients}.  

\dtr{This boundary micro\textendash scale MDEs proteolytic activity is explored via the approach initially introduced in \citet{trucu_2013_multiscale}, whereby the emergent spatio\textendash temporal dynamics of MDEs on a micro\textendash scale neighbouring envelope $\Bila(\partial\Omega(t), \epsilon/2)$ of cell\textendash scale thickness $\epsilon>0$, enabled by a bundle $\P(t)$ of overlapping cubic micro\textendash domains $\epsilon Y(z):=\Bila_{_{\lVert\cdot\rVert_{_{\infty}}}}(\zeta,\epsilon/2)$, $\forall\,\zeta\in\Omega(t)$, i.e., 
\bequd
\P(t):=\left\{\epsilon Y(z)\right\}_{_{\zeta\in\Omega(t)}}\textrm{ and } \quad\Bila(\partial\Omega(t), \epsilon/2):=\bigcup_{\epsilon Y\in \P(t)}\epsilon Y,
\eequd 
with $\Bila_{_{\lVert\cdot\rVert_{_{\infty}}}}(\zeta,\epsilon/2)$ representing the $\lVert\cdot\rVert_{_{\infty}}-$ball of radius $\epsilon/2$. This facilitates the decomposition of the overarching MDE micro\textendash process occurring on $\bigcup_{\epsilon Y \in \mathcal{P}(t)}\epsilon Y$ into an assembly of proteolytic micro\textendash dynamics occurring on each distinct $\epsilon Y$. Consequently, at any macroscopic time $t_{0}\in [0,T]$ during the tumour progression, this decomposing bundle $\P(t_{0})$ enable us to explore the MDEs micro\td dynamics on each individual micro\textendash domain $\epsilon Y\in \P(t_{0})$, where a source of MDEs emerges naturally at micro\textendash scale on the inner cancer side $\epsilon Y\cap \Omega(t)$ as result of collective contributions of the macroscopic distribution of cancer cells that arrives during the macro\textendash dynamics within a close proximity, \emph{i.e.,} within distance $\gamma_{h}>0$ from $\partial \Omega(t)$, which secretes the MDEs. Therefore, mathematically, on a small micro\textendash scale time\textendash length $\Delta t>0$ and at each micro\textendash scale spatio\textendash temporal node $(y,\tau)\in\epsilon Y\times [0,\Delta t]$, this source of MDEs induced at the micro\textendash scale by the macro\textendash dynamics is expressed through the non\textendash local term:
\bequ
\label{MDE_source}
h(y,\tau)=
\left\{
\begin{array}{ll}
 \frac{\int\limits_{\Bila_{_{\lVert\cdot\rVert_{_{\infty}}}}(y,\gamma_h)\cap \Omega(t_0)}c(x,t_0+\tau)dy}{\lambda(\mathbf{B}(y,\gamma_h)\cap \Omega(t_0))}, &\quad y\in \epsilon Y \cap \Omega(t_0), \\[0.4cm]
 0, &\quad y \notin \epsilon Y \setminus (\Omega(t_0)+\left\{z \in Y \middle|\  \Vert z\Vert_2<\rho \right\}),
 \end{array}
 \right.
\eequ
where $0<\rho<\gamma_h$ is a small mollification range, $\mathbf{B}(y,\gamma_h)$ represents the $\Vert \cdot \Vert_\infty$ ball of radius $\gamma_h$ which is centred at a micro\textendash node $y\in \epsilon Y$. Furthermore, in the presence of this source of MDEs on each of the micro\textendash domains $\epsilon Y\in \P(t_{0})$, the MDEs molecular mass\textendash transport across the tumour interface takes place on each $\epsilon Y$. Thus, denoting the MDEs density with $m(y,\tau)$, $\forall \, (y,\tau)\in \epsilon Y \times [0, \Delta t]$, this MDEs transport is assumed here to have a diffusive character and is expressed mathematically as
\bequ
\label{MDE_distribution}
 \frac{\partial m}{\partial \tau} = D_m \Delta m+h(y,\tau),\qquad \textrm{on } \epsilon Y \times [0, \Delta t], \\ [0.2cm]
\eequ
with $D_m>0$ being a constant diffusion coefficient of the MDEs, while this diffusion process is assumed to take place with: (1) null initial conditions, as this is considered to occur with \emph{no molecular memory}; and (2) with flux zero boundary conditions as we assume no MDEs molecular transport across the boundary of $\partial \epsilon Y$.} \\ 

Finally, \dtr{as this source is induced and determined directly by} the macro\textendash scale cancer cell population $c(\cdot,\cdot)$, \dtr{this} gives rise to a \emph{top\textendash down} link from the macro\textendash scale to the MDE micro\textendash scale \dtr{dynamics}. 
%Notably, this source term \eqref{MDE_source} facilitates the portrayal of the spatio\textendash temporal progression of the MDEs' micro\textendash scale distribution $m(\cdot,\cdot)$
\dtr{On the other hand, as detailed in \citet{trucu_2013_multiscale}, the pattern of peritumoural ECM degradation that the MDEs micro\textendash dynamics cause at micro\textendash scale on each boundary micro\textendash domain $\epsilon Y\in \P(t_{0})$ determines the direction of tumour boundary relocation and enables to characterise this macro\textendash scale movement of the cancer interface through rigorously derived movement laws that specifies precisely at each $x\in \partial \Omega(t_{0})$ the associated relocation direction and magnitude.} This ultimately \dtr{results in} a new evolved tumour macro\textendash domain $\Omega(t_0+\Delta t)$, \dtr{and this way a \emph{bottom\textendash up} link is established between the boundary MDEs micro\td dynamics and the macro\td dynamics}.
 
\begin{figure} [ht!]
\centering
\includegraphics[scale=0.55]{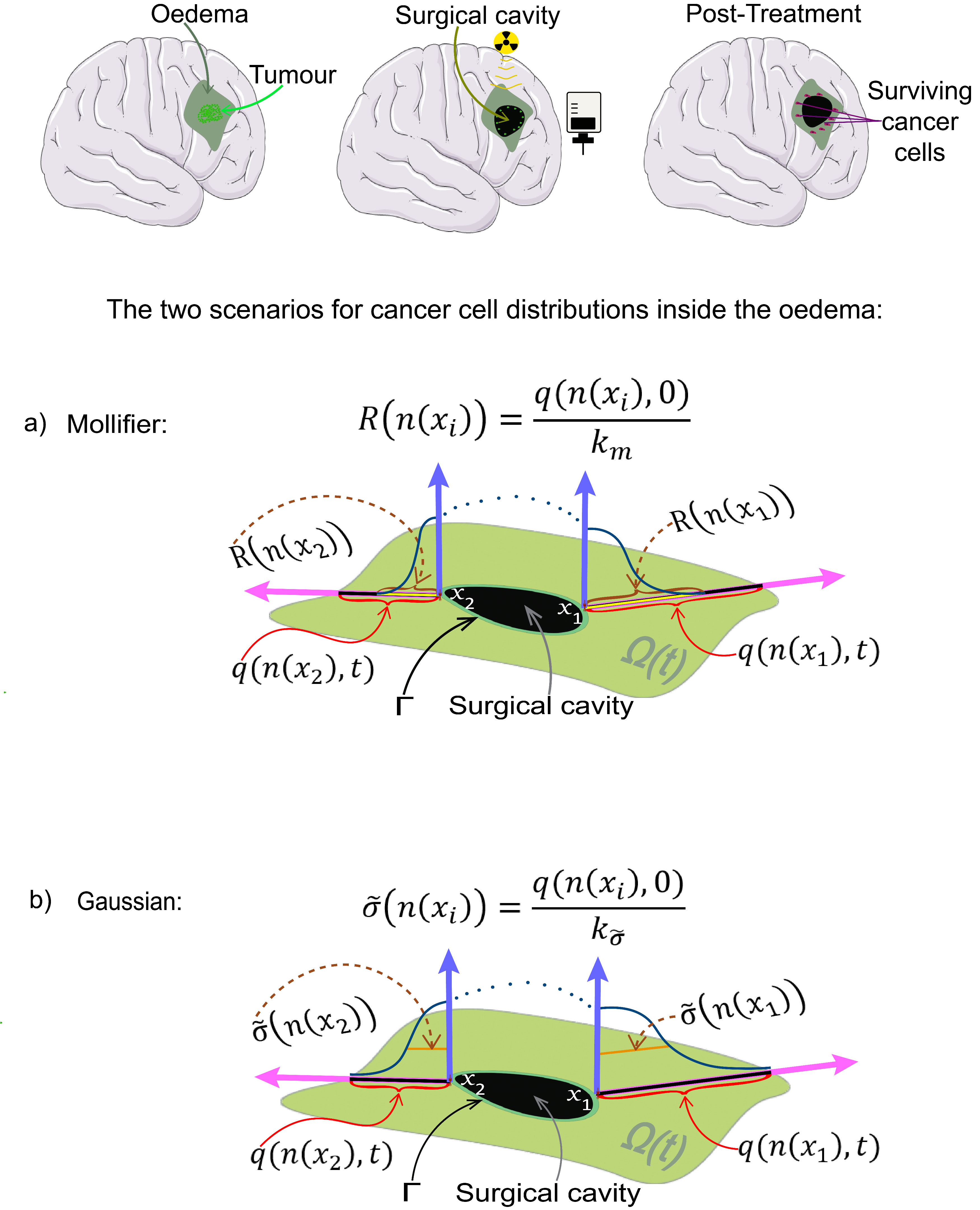}
\caption{Schematic \dg{showing} from top right to bottom left: a GBM tumour, radio and chemotherapy being applied to the surgically removed area, the surviving cancer cells inside the oedema and finally the two scenarios of cancer cell distributions used in the simulations, where $\Gamma$ represents the edge of the surgical cavity.}
\label{Schematic_1}
\end{figure}
%%%%%%%%%%%%%%%%%%%%%%%%%%%%%%%%%%%%%%%%%%%%%%%%%%%
\subsection{Reconstruction of the \am{Cancer\textendash Cell} Distribution within the Oedema} \label{Reconstruction}
\dt{It has been demonstrated that GBM cells invade the surrounding tissue via the peritumoural oedema \dg{which} is populated by phenotypically distinct cancer cells that persist in the area following surgical intervention \citep{Heterogeneity_pte}. While these cells typically remain untreated or survive the chemoradiotherapy treatment, these are not detectable on MRI scans, and contribute \dg{to} tumour \dg{recurrence}. Thus, to gain a better understanding of the tumour relapse process after surgery, \kh{it is} of interest to explore whether there is any correlation between the shape of the distribution of GBM cells that remain within oedema right after surgery and the extent of the subsequent tumour relapse. Several numerical experiments that we carried out (as those shown in Figures \am{\ref{Results MollifierDist} and \ref{Results GaussianDist})} suggest the following hypothesis, namely: 
%%%%%%%%%%%%%%%%%%%%%%%%%%%%%%%%%%%%%%%%%%%%%
%%%%%%%%%%%%%%%%Proposed Hypothesis
\begin{itemize}
\item[\textbf{H:}] \textit{a distribution of GBM cells within the oedema that has most cells mass concentrated within the immediate proximity of the cavity edge leads to a more limited spread and a slower progression of the tumour relapse.} 
\end{itemize}
This hypothesis \dt{also} aligns with clinical findings suggesting that surgical resection removes a substantial number of cancer cells, leaving the remaining cells more dispersed throughout the oedema \citep{recurrence2011patterns,petrecca2013failure}.}

	\dt{In the following, hypothesis \textbf{H} will be examined on two relevant oedema cancer cell distribution types. Furthermore, in both cases, we propose a clinical data assimilation approach, by which we aim to reconstruct the particular shape of the cancer cell distribution that enables the predictive computational modelling solution for the \am{post\textendash surgery} GBM relapse to match the available MRI imaging data.}\\

%%%%%%%%%%%%%%%%%%%%%%%%%%%%%%%%%%%%%%%%%%%%%%%%%%%
{\bf \emph{Two possible \am{post\textendash surgery} oedema cancer cell distribution scenarios:}} 
\dt{In the following, we explore hypothesis \textbf{H} by considering two possible scenarios for the \am{post\textendash surgery} oedema cancer cell spatial distribution, namely one that is compactly supported strictly within $\Omega(0)$ and one that carries \am{non\textendash zero} cell mass density distributed at any point in $\Omega(0)$. Specifically, denoting by $n(x)$ the usual outward unit normal vector to the surgical cavity edge $\Gamma$, $\forall\, x \in \Gamma$, we assume that: 
\vspace{-0.3cm}\begin{align*}
\textrm{on the positive side of the normal direction} & \textrm{ associated to any $x\in \Gamma$, represented here parametrically by}\\
& d_{_{x}}:=x+\upsilon n(x), \qquad \upsilon\geq 0,
\end{align*} 
the shape of immediate \am{post\textendash surgery} cancer cell distribution remaining within the oedema along $d_{_{x}}$, denoted here by $c_{oedema}^{d_{_{x}}}$, is of either of the following two types:\\ 
\begin{tabular}{ll}
&\\
{\textbf case 1:} & a smooth compact support \am{mollifier\textendash type} distribution of support radius $R(n(x))$ centred at $x$,\\
& which is given by\\
&\hspace{3.50cm}$
c_{oedema}^{d_{_{x}}}(v):=R(n(x), k_{_{R}})^{-1}\psi_{_{1}}\big(\frac{v}{R(n(x), k_{_{R}})}\big), \qquad v\in[0,q(n(x),0)], 
$
\\
& where $\psi_{_{1}}(\cdot)$ is the 1D standard symmetric mollifier given in Appendix \ref{mollifierAppendix}, while, for any $t\geq 0$,\\
& $q(n(x),t\!)$ denotes the distance along line $d_{_{x}}$ between $\Gamma$ and $\partial \Omega(t)$, with $R(n(x),k_{_{R}}\!)\!\!:=\!\frac{q(n(x),0\!)}{k_{_{R}}}$\\
& while $k_{_{R}}>1$ represents an uniform scaling constant applied at each $x\in \Gamma$ controls the \\
& cancer cells distribution spread in the normal direction described by $n(x)$, see Figure \ref{Schematic_1} a);\\
&\\
{\textbf case 2:} & a Gaussian distribution centred at $x$ and of standard deviation $\tilde{\sigma}(n(x))$, which is given by\\
&$\hspace{3.5cm} c_{oedema}^{d_{_{x}}}(v)\propto \mathcal{N}_{d_{x}}(0, \tilde{\sigma}(n(x), k_{_{\tilde{\sigma}}})), \qquad v\in[0,q(n(x),0)]$,\\
& where by $\mathcal{N}_{d_{x}}(0, \tilde{\sigma}(n(x),k_{_{\tilde{\sigma}}}))$ we denote here the family of normal distributions along $d_{_{x}}$,\\ 
& with $\tilde{\sigma}(n(x),k_{_{\tilde{\sigma}}}):=\frac{q(n(x),0\!)}{k_{_{\tilde{\sigma}}}}$, while $k_{_{\tilde{\sigma}}}>1$ represents an uniform scaling constant applied at\\
& each $x\in \Gamma$ controls the standard deviation, see Figure \ref{Schematic_1} b). 
\end{tabular}}	

\begin{figure}[h!]
%\begin{adjustwidth}{-1.5cm}{}
\centering
\includegraphics[width=\textwidth]{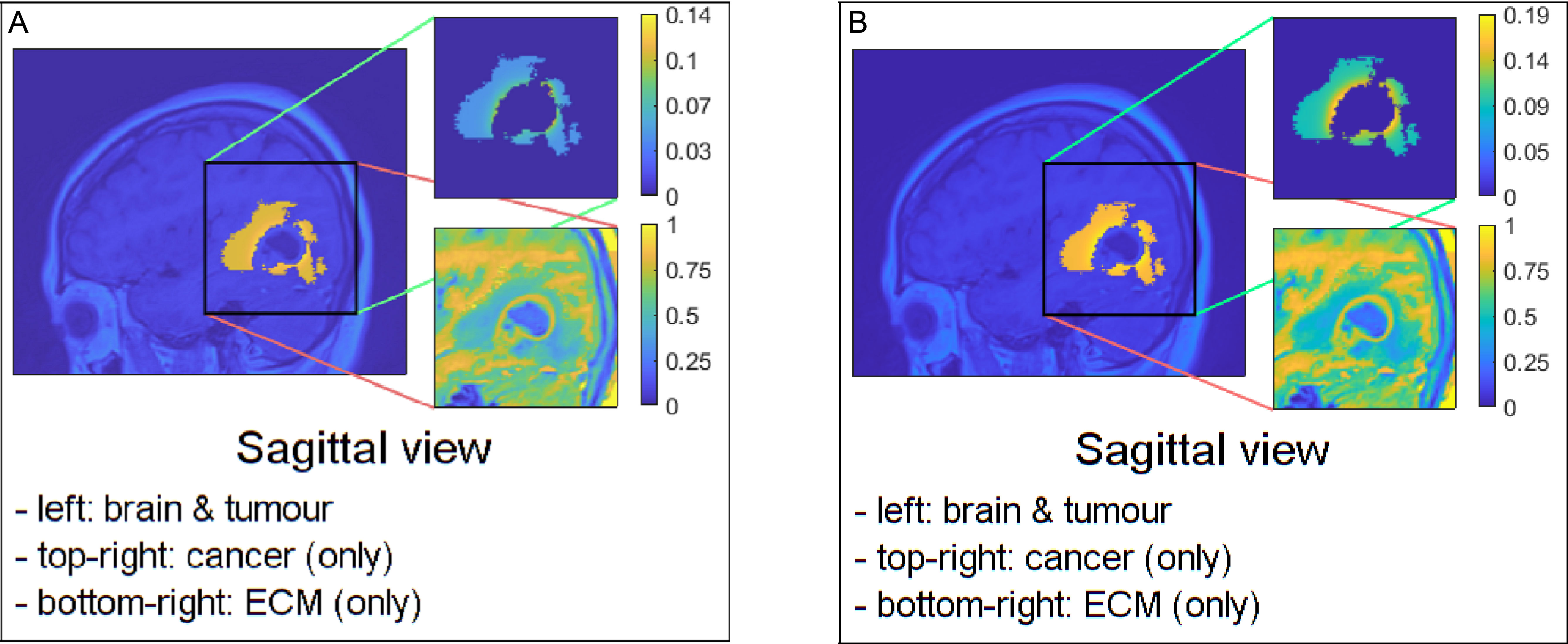} \hfill
\caption{Example of possible initial conditions when applying: \textbf{(A)} the mollifier and \textbf{(B)} the Gaussian distribution.}
\label{Initial_Conditions}
%\end{adjustwidth}
\end{figure}

\dt{For each of the two cases, we explore the correlation between the extent of significant tumour spread within oedema (characterised in case 1 by $R(n(x),k_{_{R}})$ and in case 2 by $\tilde{\sigma}(n(x),k_{_{\tilde{\sigma}}})$) and the extent of tumour invasion post\textendash surgery. A smaller $R(n(x),k_{_{R}})$ and $\tilde{\sigma}(n(x),k_{_{\tilde{\sigma}}})$} corresponds to a higher concentration of cells near the cavity's edge, with density decreasing as we move further away from it, as evident in Figure \ref{Schematic_1} and the upper\textendash right image of Figure \ref{Initial_Conditions}. \dt{Finally, we take advantage of available MRI scans to identify} suitable values for \am{$R(n(x),k_{_{R}})$ and $\tilde{\sigma}(n(x),k_{_{\tilde{\sigma}}})$} \dt{that enable the closest possible match between the computed solutions and the imaging data.} 
 
{\bf \emph{Reformulation as least square minimisation problem}:} 
\dt{In order to assimilate available MRI data to identify appropriate values of parameters controlling the degree of spread of the residual cancer cells within oedema (namely, $R(n(x),k_{_{R}})$ and $\tilde{\sigma}(n(x),k_{_{\tilde{\sigma}}})$ for case 1 and case 2, respectively), we proceed by \dg{conceptualising} this as a minimisation problem. Indeed, to achieve this, to address simultaneously both cases, we consider the mapping $Z(n(x),\cdot):(1,\infty)\to (0, q(n(x),0))$ that is defined at each $\xi\in (1,\infty)$ by
\bequ
Z(n(x),\xi):= 
\left\{
\begin{aligned}
R(n(x),\xi),& \quad \textrm{for case 1}\\
\tilde{\sigma}(n(x),\xi), & \quad \textrm{for case 2} 
\end{aligned}
\right.
\eequ
with $R(n(x),\xi):=\frac{q(n(x),0)}{\xi}$ in case 1, and $\tilde{\sigma}(n(x),\xi):=\frac{q(n(x),0)}{\xi}$ in case 2. In this context we aim to identify the point of minimum $\xi_{min}$ (representing the optimal controller parameters  $\bar k_{_{R}}$ and $\bar k_{_{\tilde{\sigma}}}$) in case 1 and case 2, respectively) that minimises the following $\xi-$dependant distance 
\bequ \label{minEq}
dist (c_{_{Z(n(x),\xi)}}, MRI):=\max_{i=1,...,N_{_{data}}}\Vert c_{_{Z(n(x),\xi)}}(\cdot,t_{_{i}}) - MRI_{_{i}} \Vert_{_{2}},
\eequ
%and 
%\bequ
%dist (c_{_{\tilde{\sigma}(n(x),\cdot)}}, MRI_{_{relapse}}):=\Vert c_{_{\tilde{\sigma}(n(x),\cdot)}} - MRI_{_{relapse}} \Vert_{_{2}},
%\eequ
where $\{t_{_{i}}\}_{_{i=1,...,N_{_{data}}}}$ are the macroscopic times at which the corresponding $MRI$ scans $\{MRI_{_{i}}\}_{_{i=1,...,N_{_{data}}}}$ will have been recorded. Here, $c_{_{Z(n(x),\xi)}}(\cdot, t_{_{i}})$ represents the spatial density of the computed solution evaluated at $t_{_{i}}$ that is obtained for a \emph{guessed initial condition} $c^{guess}_{0}(\xi;d_{_{x}}, v)$ that corresponds to $\xi\in(1,\infty)$. Finally, for each $\xi\in (1,\infty)$, the guessed initial condition $c^{guess}_{0}(\xi;d_{_{x}}, v) $ is defined in each of the two cases as:\\
 \begin{tabular}{lll}
 &&\\
{\textbf case 1:} & $c^{guess}_{0}(\xi;d_{_{x}}, v):= R(n(x), \xi)^{-1}\psi_{_{1}}\big(\frac{v}{R(n(x), \xi)}\big)$, & $\quad v\in[0,q(n(x),0)]$,\\
&&\\
{\textbf case 2:} & $c^{guess}_{0}(\xi;d_{_{x}}, v)\propto \mathcal{N}_{d_{x}}(0, \tilde{\sigma}(n(x), \xi))$, & $\quad v\in[0,q(n(x),0)]$.
\end{tabular}}
\subsection{Clinical Data Assimilation}

\subsubsection{\dt{Acquisition of} Clinical Data}

The clinical data used for this study was acquired from one out of 48 GBM patients who received different treatments at Ninewells Hospital between 2017 and 2021, \kh{chosen due to their prolonged survival, giving us access to multiple MRI scans which can be used to improve our mathematical model}. \amr{Ethical approval was obtained from the local Caldicott Guardian, Integrated Research Application System (IRAS)(project ID: 309957), Tayside Research and Development Committee (project ID: 2022NH01) and Research Ethics Committee (REC) (Ref: 22/NS/0021)}. To be included in the study, patients had to be over 16 years old but no older than 85 years old, with histologically confirmed GBM, and have undergone multiple pre\textendash operative and post\textendash operative MRI scans and received standard NHS chemotherapy and radiotherapy treatments. Patients with a limited number of MRI scans were excluded.

\subsubsection{Brain Imaging, Preprocessing and Segmentation.} \label{StagesExplanation}

The MRI scans were conducted using \dg{NHS} GE 1.5 Tesla scanners and included multiple pre\textendash operative and post\textendash operative scans for the selected patient. The scans consisted of T1\textendash weighted (T1), T2\textendash weighted (T2), contrast\textendash enhanced T1\textendash weighted (with Gadolinium) (T1+C), diffusion\textendash weighted imaging (DWI) (for specific dates), and T2\textendash FLAIR sequences.

\begin{figure}[ht!]
\centering
{\includegraphics[width=\textwidth]{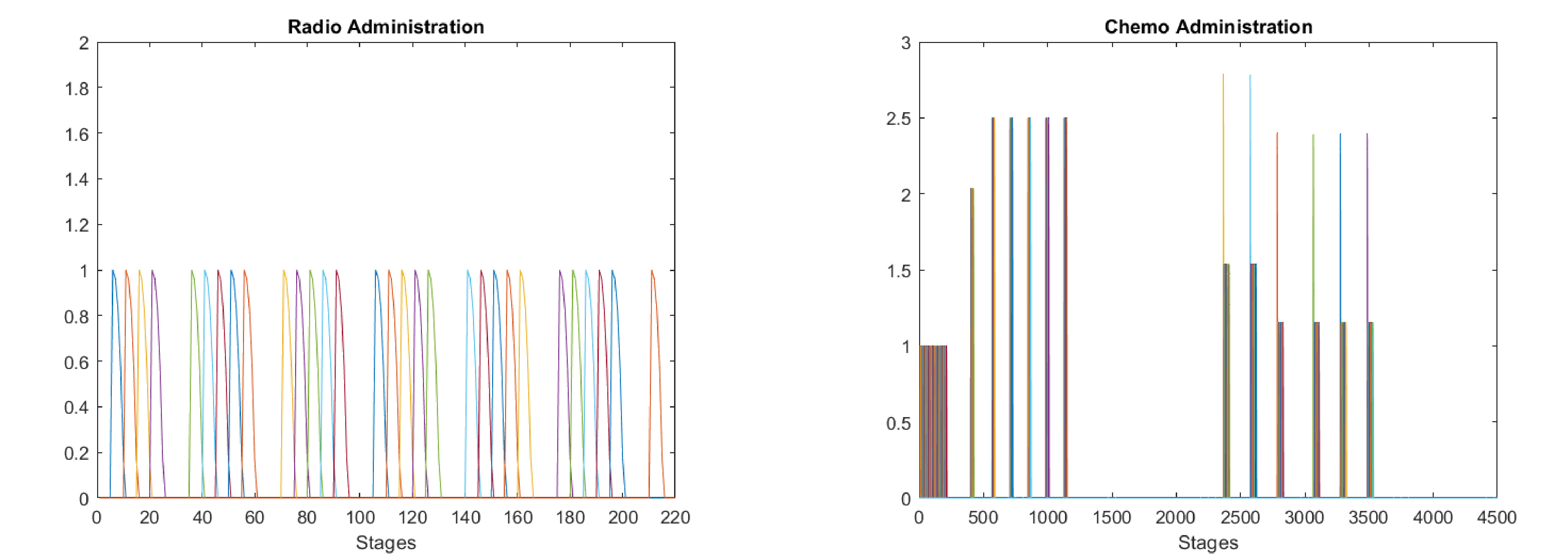}}
\caption{\kh{Visualizing treatment dynamics: These graphs depict the radiotherapy and chemotherapy delivery for this patient. The horizontal axis represents the computational stages of the treatment simulation, with every five stages corresponding to one actual day. The vertical axis represents the intensity of the radiation and chemotherapy doses delivered at each stage.}}
\label{TherapyAdmins}
\end{figure}

A single typical patient from the series was used for the calculations described here. The patient received initial surgery, followed by chemoradiotherapy with Temozolomide (TMZ) at 130 mg per day concurrently with radiotherapy at a total of 60 Gy distributed equally in 30 total fractions, following the Stupp protocol, and in addition, adjuvant TMZ at a dose of 265\textendash 325 mg (6 cycles) after the initial radiotherapy and chemotherapy treatments. Due to recurrence, \kh{visible seven months after the completion of concurrent chemoradiotherapy and adjuvant TMZ}, the patient also received Lomustine at 160 mg, Procarbazine at 150\textendash 200 mg, and Vincristine at 1\textendash 2 mg (6 cycles). The delivery of the radiotherapy and chemotherapy can be seen in Figure \ref{TherapyAdmins}, \kh{where we considered all chemotherapeutic drugs functionally equivalent (TMZ, Lomustine, Procarbazine and Vincristine), adjusting only the dosage based on the treatment plan.} \am{With the purpose to simulate this treatment delivery, we need to modify the model such that every \kh{computational} macro\textendash micro stage corresponds to a certain amount of real time. This is later explained in section \ref{TreatmentMRI}.} 

	The patient underwent two more surgeries, and MRI scans were taken before and after each of the surgeries, as well as after the completion of the different treatments. When the patient was not undergoing any treatments, MRI scans were conducted every three to four months.
	\\The MRI scans were first pre\textendash processed using Statistical Parametric Mapping (SPM\textendash 12, http://www.fil.ion.ucl.ac.uk/spm/). This pre\textendash processing involved reslicing, normalising and finally segmentation of the T1 scan to obtain the white and grey matter densities. As Diffusion Tensor Imaging (DTI) scans were not obtained for this patient, we modified a standard DTI scan from a healthy volunteer from the IXI Dataset (http://brain-development.org/ixi-dataset/), which was warped to match the anatomy of the T1 scan of the GBM patient, hence we were able to infer brain fibre tract directions for the GBM brain.
	
\begin{figure}[ht!]
\centering
{\includegraphics[width=\textwidth]{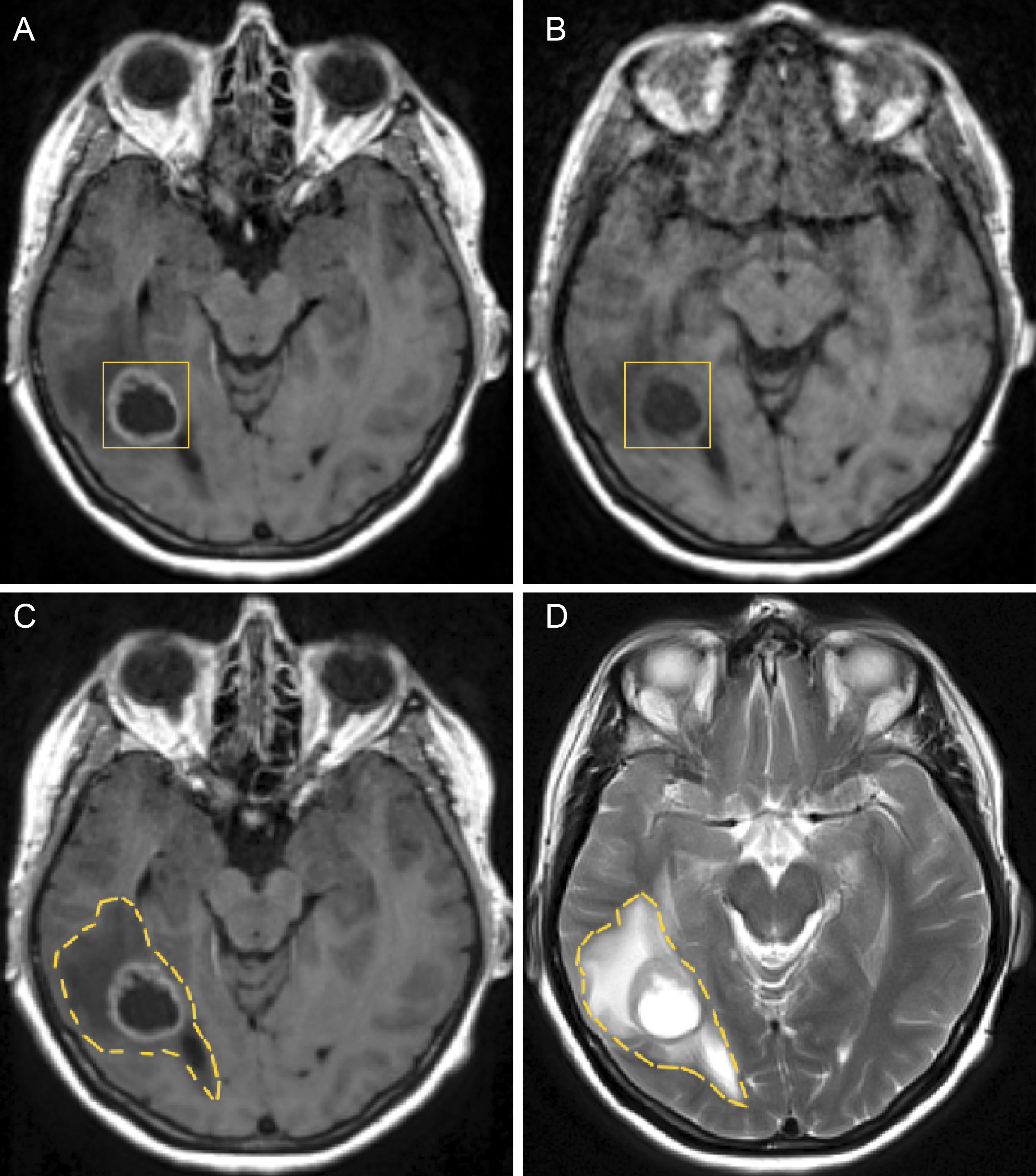}}
\caption{Figure showing GBM and oedema in different \kh{axial} MRI scans: \textbf{(A)} GBM tumour in a T1+C, \textbf{(B)} GBM tumour in a T1 without contrast, \textbf{(C)} oedema in a T1+C and \textbf{(D)} oedema in a T2 \dg{weighted scan.}}
\label{GBM_Oedema_Scans}
\end{figure}

	In Figure \ref{GBM_Oedema_Scans}, on the top right, a T1\textendash weighted scan (T1) is shown and on the top left a T1 scan with gadolinium contrast (T1+C), which outlines the tumour as gadolinium is taken up by the invasive edge of the tumour.	 The GBM proliferating edge is observed as enhancing in the T1+C and hypo- to iso- intense in T1, as seen in Figure \ref{GBM_Oedema_Scans}. The region which is hyperintense in the T2 scan (and T2\textendash FLAIR) and it is non\textendash enhancing in T1+gadolinium represents the oedema, as seen in the bottom row of Figure \ref{GBM_Oedema_Scans}.
	
\begin{figure}[ht!]
\centering
{\includegraphics[width=\textwidth]{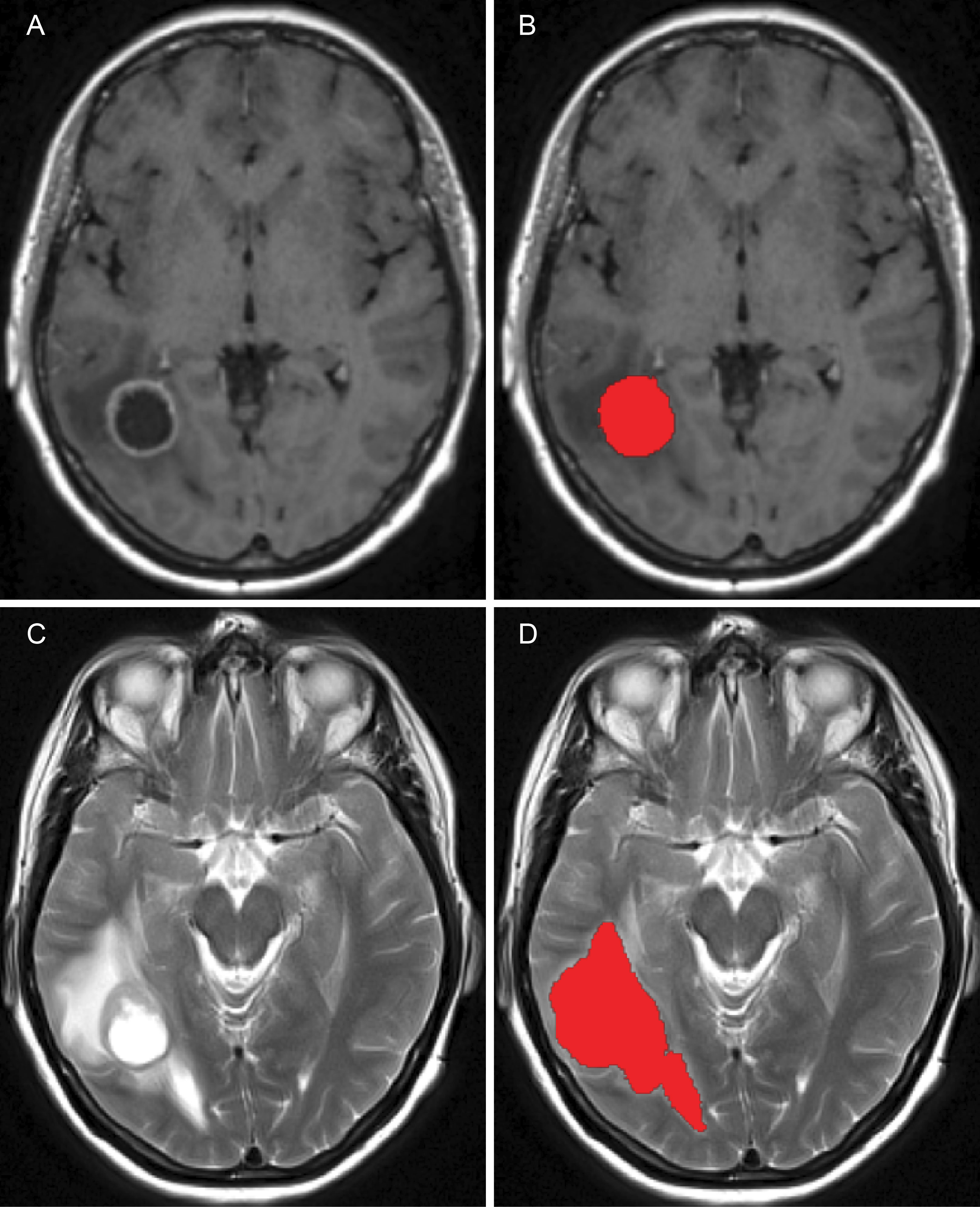}}
\caption{Figure \dg{showing} the \kh{axial MRI} scans and corresponding volumes of interest for the patient in question: \textbf{(A)} T1+C scan, \textbf{(B)} Pre\textendash surgical GBM VOI, \textbf{(C)} T2 scan and \textbf{(D)} oedema VOI.}
\label{GBMinScans}
\end{figure}

	After pre\textendash processing has been completed, tumour segmentation was done using MRIcroGL, version v1.2.20220720 (www.nitrc.org). The segmentation was performed manually, under the supervision of NHS Consultant neurosurgeons \mo{KHI} and \mo{MO}, who specialise in the treatment of GBM. The scans were processed on a axial  (transverse) slice\textendash by\textendash slice basis, as seen in \am{Figure \ref{GBMinScans}}, for the post\textendash contrast T1 and T2 sequences. \dt{These enabled the exploration of important characteristics, referred to as \emph{``volumes of interest"} (VOI), \amr{one for the pre\td surgical tumour and another one for the oedema before surgery}, which were given in the form of binary masks (\emph{i.e.,} individual indicator matrices of zeros and ones that give the footprints of the tumour and oedema) and that} were later used in our mathematical model.
	
%
%	We will seek to identify this, relative to the data provided by our measurements so that the discrepancies between the spread recorded and the measurements provided are minimised. Mathematically, it means:
%	\begin{equation*}
%	\sigma= \operatorname{argmin}\{\Vert c(x,t) -  measurments \Vert_2\}
%	\end{equation*}
%	
%	Finally, we assume that outside of the oedema boundary, the cancer density is zero. 

\section{Results}

\dt{T}he numerical \dt{approach} employed in this work to tackle both the macro\textendash scale and micro\textendash scale dynamics, \dt{as well as the \am{top\textendash down} and bottom up links between the scales, builds on a sequence of multiscale modelling and computational works \amr{introduced} in \citet{trucu_2013_multiscale,shuttleworth_2019_multiscale,shuttleworth_2020,suveges_2021_mathematical,Szabolcs_2022_Nutrients}, and extends these through the introduction of a new governing equation for capturing nutrients dynamics. Moreover, to identify the shape of the remaining \am{post\textendash surgery} oedema cancer cell population distribution that lead to GBM relapse, the 3D computational modelling platform developed here is coupled with a \am{least\textendash square\textendash type} clinical data assimilation approach using \am{post\textendash surgical} MRI scans. } 

	Similar to the methodology outlined in \citet{Szabolcs_2022_Nutrients}, we utilise the successive over\textendash relaxation method for solving the nutrients Equation \eqref{Nutrient_Eq}. For the rest of the macro\textendash scale \dt{dynamics in \eqref{CompleteMacro}, we follow similar steps as in \citet{suveges_2021_mathematical,Szabolcs_2022_Nutrients} and employ the method of lines with the following \dg{details}. Specifically, the spatial operators (\emph{i.e., } the diffusion and adhesion operators) are addressed as follows: (a) for diffusion we implement a symmetric finite difference scheme based on convolution, as detailed in \citet{Szabolcs_2022_Nutrients}; and (b) for adhesion we utilise a convolution\textendash driven approach employing a \textit{fifth\textendash order weighted essentially non\textendash oscillatory} (WENO5) finite difference scheme \citep{Liu_1994,Jiang_1996,Zhang_2006,Kim_2005}, also elaborated upon in \citet{Szabolcs_2022_Nutrients}. Finally, the time marching is ensured through a predictor corrector scheme introduced in \citet{shuttleworth_2019_multiscale} and further detailed in \citet{suveges_2021_mathematical,Szabolcs_2022_Nutrients}.}
	
%	for spatial discretisation of the system \eqref{CompleteMacro}. Subsequently, we apply a non\textendash local predictor\textendash corrector scheme to address the resulting ODEs \citep{suveges_2021_mathematical,Szabolcs_2022_Nutrients}.

%%	In our approach, we achieve precise approximations of the spatial operators, specifically the diffusion and adhesion operators, through efficient convolution\textendash based approaches. For diffusion, we implement a second\textendash order central difference scheme based on convolution, as detailed in \citet{Szabolcs_2022_Nutrients}. Meanwhile, for adhesion operators, we utilise a convolution\textendash driven approach employing a \textit{fifth\textendash order weighted essentially non\textendash oscillatory} (WENO5) finite difference scheme, also elaborated upon in \citep{Szabolcs_2022_Nutrients}.

%	For more in\textendash depth information, we recommend consulting the references \citep{suveges_2021_mathematical, Szabolcs_2022_Nutrients}.
	 
\subsection{Treatment scheduling}\label{TreatmentMRI}

One of the primary aims of \dg{our work} is to accurately replicate the treatment regimen and dosages administered to a specific patient, \amr{which in this case,} revolves around the time span bridging the \am{first and second surgery, during which various treatment modalities were employed throughout this entire duration.}

The comprehensive timeline for this patient extends beyond 900 days, encompassing the period between the first and second surgery. During this span, chemotherapy and radiotherapy were administered, and MRI scans were conducted on specific dates. In order to forecast the possibility of relapse and tumour spread based on this patient's treatment timeline, we need to simulate the treatment process over the course of these \am{900\textendash plus} days. To achieve this goal, it is important to demarcate the \kh{computational} \am{macro\textendash micro} stages and steps meticulously. 
\\\am{\kh{To precisely capture the daily dynamics of the patient's treatment, the model was modified with a temporal discretisation scheme. Here, every five computational stages represent one actual day, resulting in over 4500 stages}, as shown in Figure \ref{TherapyAdmins}. This discretisation comes from splitting the macro\textendash scale time interval (\kh{\emph{i.e.,} the treatment duration}) $[0, T_{f}]$ into smaller intervals $\{[k\Delta t, (k+1)\Delta t]\}_{k=0,k_{_{max}}}$. Each such increment,which encompasses both the macro\textendash dynamics that takes place on $\Omega(k\Delta t)$ over the time period $[k\Delta t, (k+1)\Delta t]$ and the micro\textendash dynamics at its boundary (influenced by the ``top\textendash down" links (explained previously in greater detail in section \ref{Links}) on each of the boundary micro\textendash domains $\epsilon Y\in \mathbf{B}(\partial \Omega(k\Delta t), \epsilon/2)$) constitutes a \emph{``stage k"}. As described in \citet{trucu_2013_multiscale,alzahrani_2019_multiscale}, these micro\textendash dynamics at the boundary dictate the precise direction and displacement magnitude for the relocation of each of the points on $\partial \Omega(k\Delta t) $), progressing this way the stage k tumour domain $ \Omega(k\Delta t) $ into the newly obtained domain $\Omega((k+1)\Delta t)$. With this method, we can match the exact treatment for each day and compare our simulations with the MRI scans taken on those specific dates.}

\subsection{Initial conditions}

The initial micro\textendash fibre distribution within a micro\textendash domain $\delta Y(x)$ is \dt{considered here to be the one introduced in \citet{suveges_2021_mathematical}, which in brief can be summarised as follows. On one hand, if $x\in Y$ is located in the grey matter zone, random straight narrow 3D\td stripes (\emph{i.e.,} narrow equal\td square cross\td section parallelepipedic bars that fit within $\delta Y(x)$) are distributed until the ratio of the cumulative stripe volume occupied $35\%$ out of the entire $\delta Y(x)$}. On the other hand, if $x$ is located in the white matter, a predefined set of aligned \dt{straight narrow 3D\td stripes is distributed within $\delta Y(x)$ until the volume is filled up to the same percentage, \emph{i.e., } up to $35\%$}. We also incorporated information about the white and grey matter tracts from the T1+C scan into the micro\textendash scale fibre distribution \citep{suveges_2021_mathematical}.
For the non\textendash fibre ECM phase, we have the following initial condition:
\begin{equation}
l(x,0) = \min\{h(x_1,x_2,x_3),1-c(x,0)\},
\end{equation}
where for any $x:=(x_1,x_2,x_3) \in Y$ we have:
\begin{equation*}
h(x)=\frac{1}{2}+\frac{1}{4}\sin(7\pi y_{_{1}}(x)y_{_{2}}(x)y_{_{3}}(x))^3 \cdot \sin(7\pi y_{_{1}}(x)/y_{_{2}}(x)/y_{_{3}}(x)),
\end{equation*}		
with:
\begin{equation*}
\begin{array}{lll}
y_1(x)&:=&\frac{1}{3}(x_1+1.5),\\ [0.2cm]
y_2(x)&:=&\frac{1}{3}(x_2+1.5),\\ [0.2cm]
y_3(x)&:=&\frac{1}{3}(x_3+1.5).
\end{array}
\end{equation*}
Lastly, the initial condition for the nutrients is set to: $\sigma(x,0)=0.4.$

\subsection{Numerical Simulations}

This section presents the results of 3D numerical simulations of the multiscale model of GBM tumour growth. The parameter values used in the simulations are taken from Table 0 in Appendix A. Any modifications made to the values are stated in the text. 
	
	To display the evolution of the tumours at time $45\Delta t$, we show four panels for each simulation. The first three panels show the tumour in the coronal, axial, and sagittal planes, respectively. The final panel shows a 3D image of the brain with the embedded tumour alongside the 3D tumour in isolation.
\\The figures below show the evolution of GBM tumours with different cancer cell distributions in the oedema, under the application (or not) of radiotherapy and chemotherapy. The densities of the main tumour and the ECM are shown in the top\textendash right and bottom\textendash right corners of each of the three classical\textendash views panels, respectively.
	
	\amr{Now, to initialise our simulations, we use the manually segmented masks for both the pre\td surgical oedema and tumour, which are subtracted in order to create a surgical cavity, as depicted in Figure \ref{Schematic_1}. Next, we apply either a cancer cell distribution within the modified oedema mask of the shape of a mollifier\td type distribution or a Gaussian\td type distribution.}
	%\am{As stated in Section \ref{Reconstruction}, the starting point of each of the simulations is based on the manually segmented masks of a specific patient with GBM, specifically a mask representing the oedema before surgery and another mask representing the initial tumour before surgery, which is subtracted from the oedema mask and assumed to be the surgically removed tumour, as depicted in Figure \ref{Schematic_1}.} 
Moreover, the treatment used on this specific patient is also being applied at the simulation, \amr{as shown in Figure \ref{TherapyAdmins}.}
		
\begin{figure}[ht!]
%\begin{adjustwidth}{-1.5cm}{}
\centering
\includegraphics[width=\textwidth]{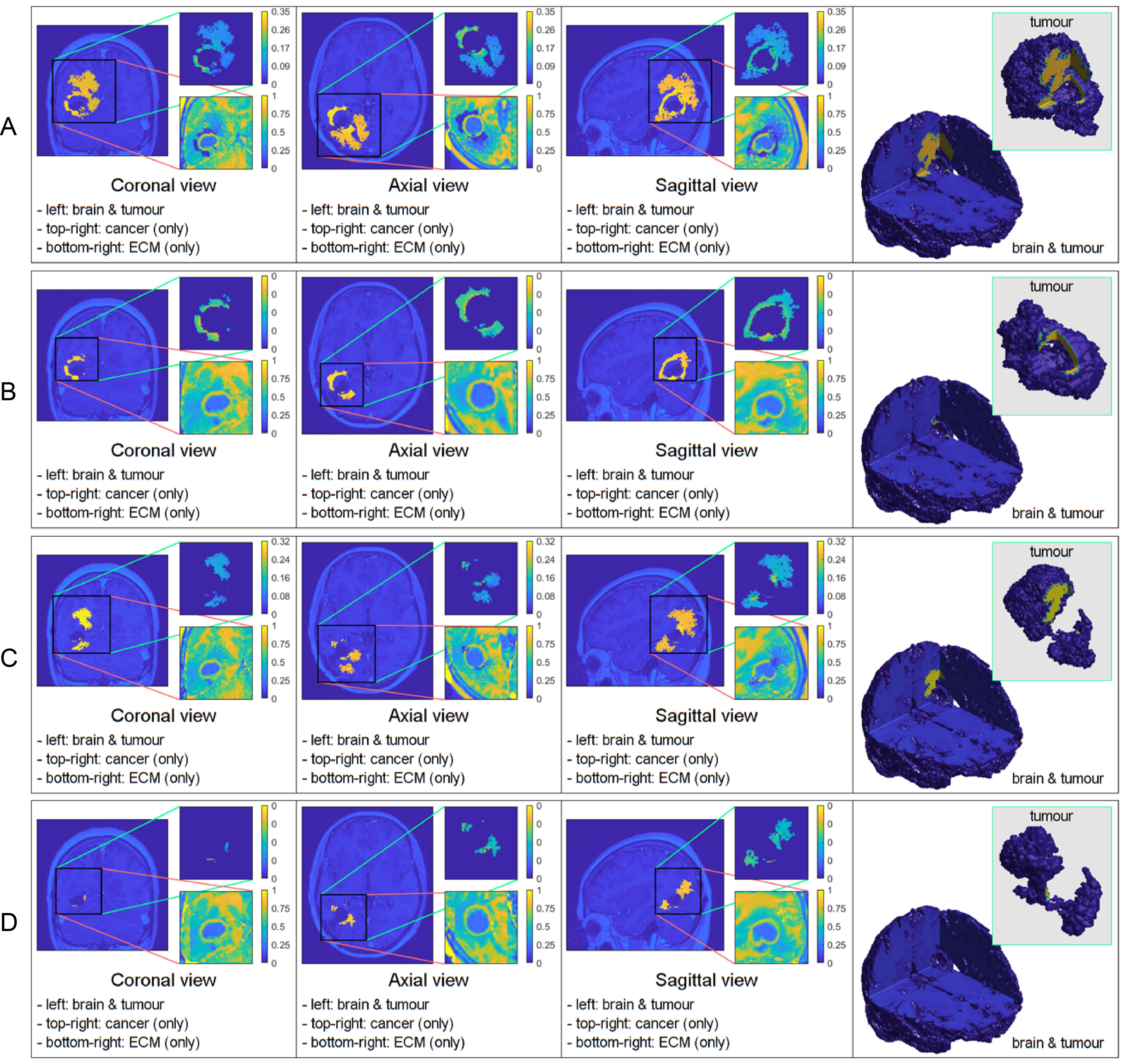} \hfill 
\caption{Comparative 3D simulations featuring the mollifier distribution: \textbf{(A)} $k_{_{R}}=5$ with no treatment; \textbf{(B)} $k_{_{R}}=5$ with treatment; \textbf{(C)} $k_{_{R}}=20$ with no treatment; and \textbf{(D)} $k_{_{R}}=20$ with treatment. All simulations captured at \am{macro\textendash micro} stage 45.}
\label{Results MollifierDist}
%\end{adjustwidth}
\end{figure}

\begin{figure}[ht!]
%\begin{adjustwidth}{-1.5cm}{}
\centering
\includegraphics[width=\textwidth]{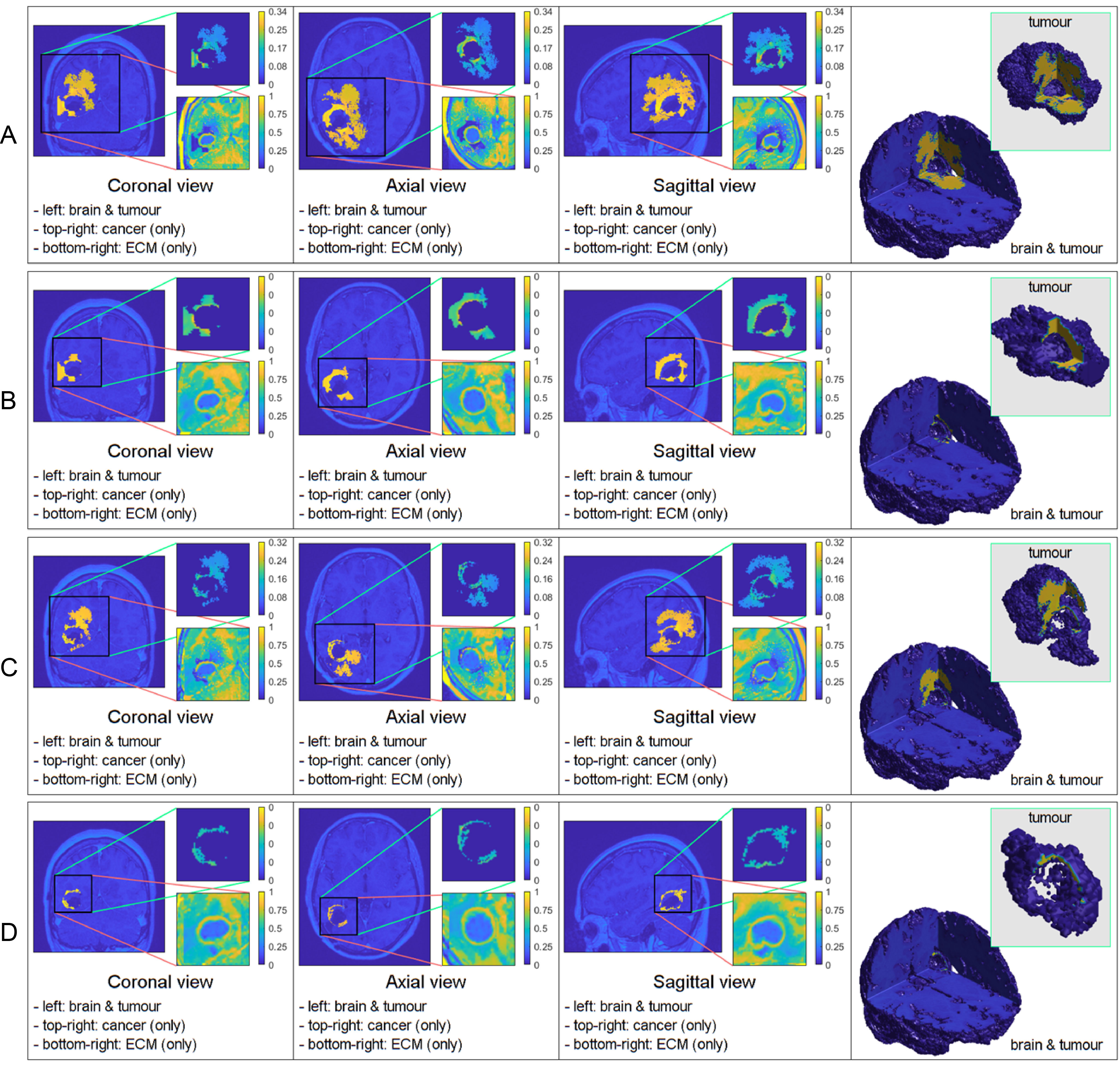} \hfill
\caption{Comparative 3D simulations featuring the Gaussian distribution: \textbf{(A)} $k_{\tilde{\sigma}=10}$ with no treatment; \textbf{(B)} $k_{\tilde{\sigma}=10}$ with treatment; \textbf{(C)} $k_{\tilde{\sigma}=100}$ with no treatment; and \textbf{(D)} $k_{\tilde{\sigma}=100}$ with treatment. All simulations captured at \am{macro\textendash micro} stage 45.}
\label{Results GaussianDist}
%\end{adjustwidth}
\end{figure}

\begin{figure}[ht!]
%\begin{adjustwidth}{-1.5cm}{}
\centering
\hfill
\includegraphics[width=\textwidth]{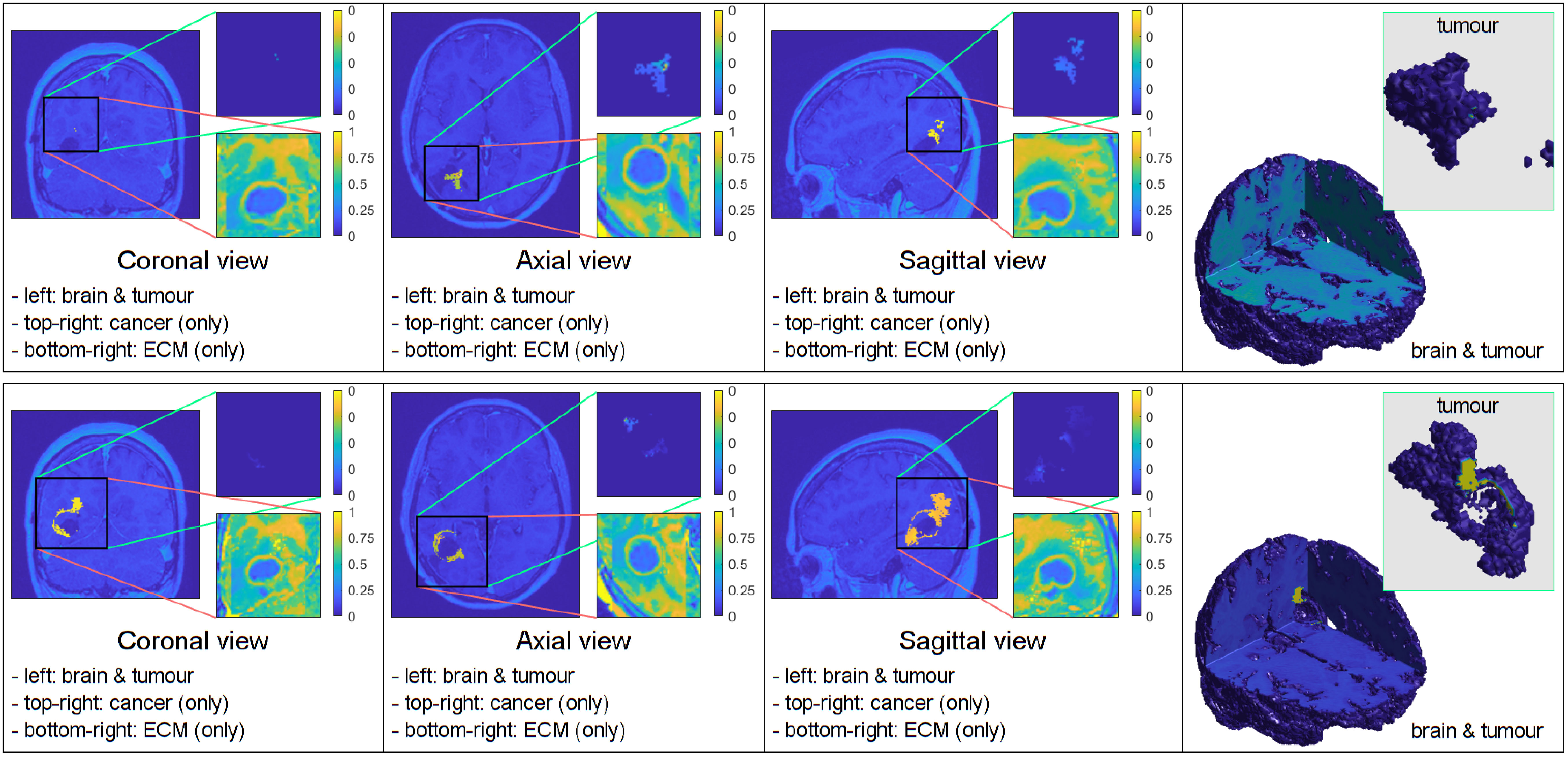} \hfill
\caption{Simulations that showed the least tumour progression with the mollifier (top) and Gaussian distribution (bottom), respectively.}
\label{BestResults}
%\end{adjustwidth}
\end{figure}

	The figures below show the results of applying the mollifier distribution with different values for \am{$k_{_{R}}$}, in Figure \ref{Results MollifierDist} and the Gaussian distribution with different \amr{values for} \am{$k_{_{\tilde{\sigma}}}$}, in Figure \ref{Results GaussianDist}. Finally, we compare the results which showed a reduction in tumour size, as shown in Figure \ref{BestResults}.

	The results of the simulations are consistent with clinical data, \amr{which} have shown that the highest concentration of cancer cells in recurrent GBM patients are located at the resection margin \citep{recurrence2011patterns,petrecca2013failure}, hence \amr{using the oedema mask, and applying either a mollifier or Gaussian distribution of cancer cells within it, can lead to clinically relevant results by adjusting $k_{_{R}}$ or $k_{_{\tilde{\sigma}}}$, respectively.}

	Figure \ref{Results MollifierDist} \amr{illustrates the results from two experiments. In the first experiment, rows A) and B), we set the parameter $k_{_{R}}=5$. Row A) depicts the results obtained without applying any treatment, whilst row B) shows the simulation when the treatment from Figure \ref{TherapyAdmins} was applied throughout the macro\td micro stages.
	The second experiment, showcased in rows C) and D), used $k_{_{R}}=20$. Similarly to the first experiment, row C) presents the results without any treatment, while row D) showcases the simulation with the treatment applied.}
	%\am{using the mollifier distribution}, in rows A) with $k_{_{R}}=5$ and C) with $k_{_{R}}=20$, whilst rows B) with $k_{_{R}}=5$ and D) with $k_{_{R}}=20$ \dg{show} the results obtained when applying the specific treatment used for the patient in question. 
	Observe that applying the treatment, Figure \ref{Results MollifierDist} rows B) and D), highly reduces the densities and spread of the tumour, but there are still residual cancer cells left, mostly around the surgical cavity. Finally, observe that increasing the \am{value of $k_{_{R}}$} leads to less spread, when comparing the top two rows (A and B) with the bottom two (C and D).  
	
Similarly to the previous case, Figure \ref{Results GaussianDist} displays the simulations \am{using the Gaussian distribution with no treatment being applied in rows A) and C), whilst rows B) and D)} are the simulations with chemoradiotherapy. Moreover, we set \am{a value of $k_{_{\tilde{\sigma}}}=10$} for both rows A) and B) and $k_{_{\tilde{\sigma}}}=100$ for both rows C) and D).
	When using the Gaussian \dt{distribution for the residual cancer cells within oedema after surgery}, we observe a similar morphology to the previous case. \amr{As with the mollifier distribution experiment,} increasing \am{$k_{_{\tilde{\sigma}}}$} and applying the treatment also leads to less tumour growth and spread. Nonetheless, \dt{this} still leads to a bigger tumour, and with much more spreading potential than \dt{in} the mollifier \dt{case}.

	Furthermore, we performed experiments with different parameter values and found that the most compact and least invasive tumour spread was obtained when applying the chemoradiotherapy treatment to an initial maximum cancer cells density of 0.1, \amr{followed by applying the mollifier distribution to it within the oedema,} with $k_{_{R}}=30$. As shown in Figure \ref{BestResults} top row this \amr{approach leads to} barely any growth, and the tumour remains stable throughout the stages. \dt{Moreover, within the same scenario but considering} the Gaussian \dt{distribution of cells within oedema} with $k_{_{\tilde{\sigma}}}=100$, showcased in Figure \ref{BestResults} bottom row, \dt{this also leads to less spread than in the previous experiment from Figure \ref{Results GaussianDist}}, but as showcased in the 3D panel of Figure \ref{BestResults} for the Gaussian distribution simulation, the tumour is larger and \dg{spreads} more \dt{than in the mollifier case from Figure \ref{BestResults}}.

\begin{figure}[ht!]
%\begin{adjustwidth}{-1.5cm}{}
\centering
\includegraphics[scale=0.5]{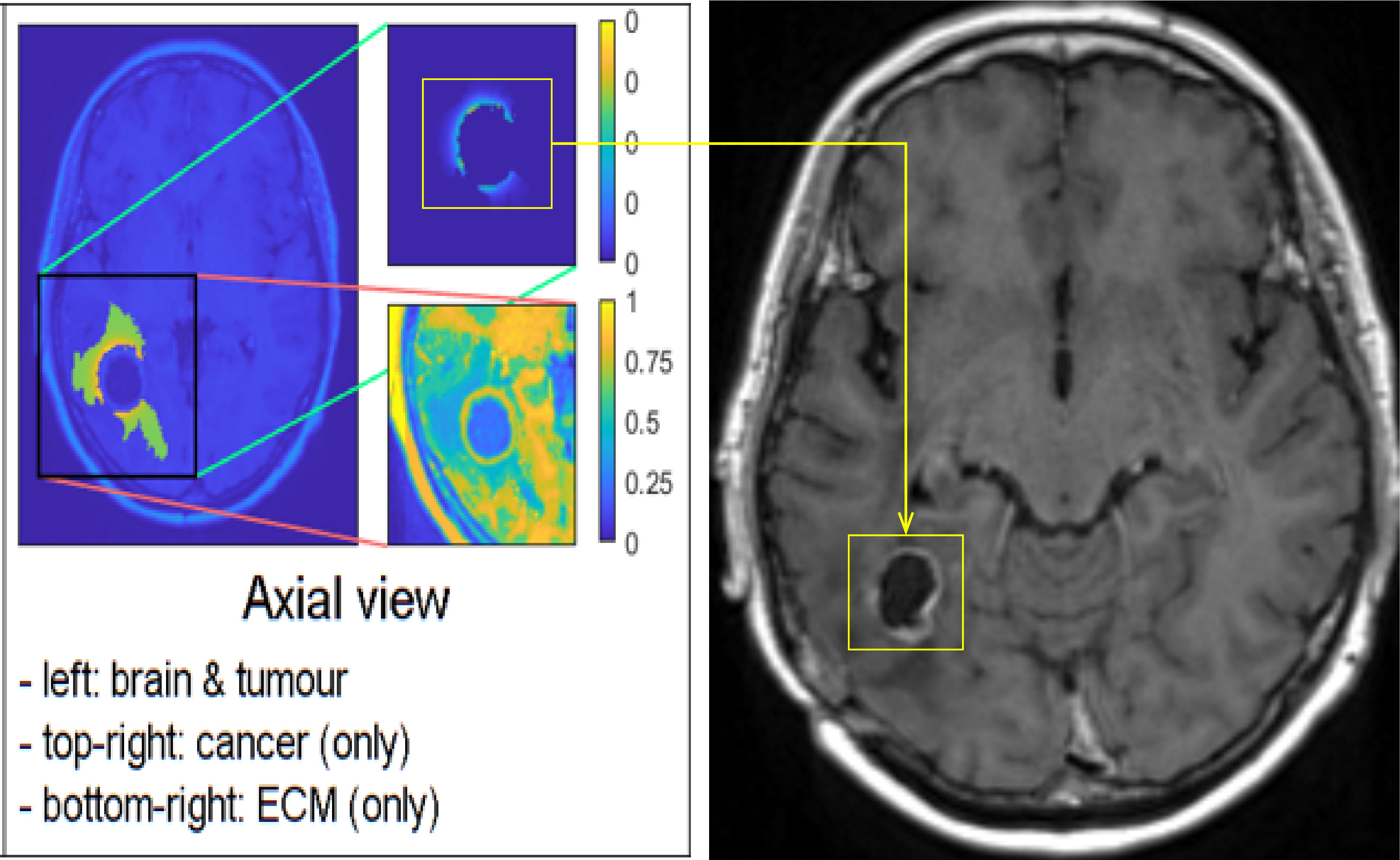} \hfill 
\caption{Visual comparison between one of our simulations and the MRI scan of the patient, taken 881 days after the first surgery.}
\label{Comparison}
%\end{adjustwidth}
\end{figure}

	Finally, during the course of various experiments, we observed an intriguing outcome. When we applied the mollifier distribution \amr{to a specific set of values,} the resulting outcome closely resembled an MRI scan taken 881 days into the patient's treatment, as evidenced by a visual comparison between the top\textendash right image of our simulation and an actual MRI scan of the patient, as shown in Figure \ref{Comparison}. This discovery guided us toward the subsequent phase of our goal: the comparative analysis of our simulations with MRI scans from this particular patient, \amr{enabled by the modification of $k_{_{R}}$ or $k_{_{\tilde{\sigma}}}$ so that our simulations can closely match the given imaging data.}

\subsection{Comparison between our simulations and the MRI data}
	
\am{A key objective of this work is to predict the growth} dynamics of GBM tumours whilst incorporating a range of pre\textendash operative and post\textendash operative MRI scans from the specific patient into our analysis. This pursuit, crucial in the field of neuro\textendash oncology, demands a thorough examination of the treatments received by these patients plus the analysis of the MRI scans. This examination is carried out through rigorous comparisons between our computational simulations and the existing MRI scans.

To initiate this process, it is essential to adjust and refine the parameters of our computational model, ensuring that the complex details match the real information found in the MRI scans.
\am{\kh{In order} to properly compare our simulations to MRI scans, we} start by aligning the simulation data with the corresponding MRI scan taken at a specific time in our timeline. Comparing the manually outlined tumour volume from the MRI, \dg{outlined under the supervision of \mo{KHI and MO}}, with our simulated cancer density, we calculate the absolute difference, \am{following the  methods described in Section \ref{Reconstruction}, such that Equation \eqref{minEq} is satisfied.} If, at any time point, the cancer growth exceeds a set threshold and the disparities between the actual and predicted data are significant, we halt the simulation. Subsequently, we adjust either $k_{_{R}}$ or $k_{_{\tilde{\sigma}}}$ in a dyadic fashion until the simulation closely matches the real data, meeting our predefined threshold.
\\This iterative refinement process ensures that our simulations accurately represent tumour dynamics observed in MRI scans, thereby enhancing the reliability and applicability of our computational models.
\subsubsection{Utilising Post\textendash Surgical MRI Scans for More Realistic Tumour Simulations}
\begin{figure}[ht!]
\centering
\includegraphics[scale=0.5]{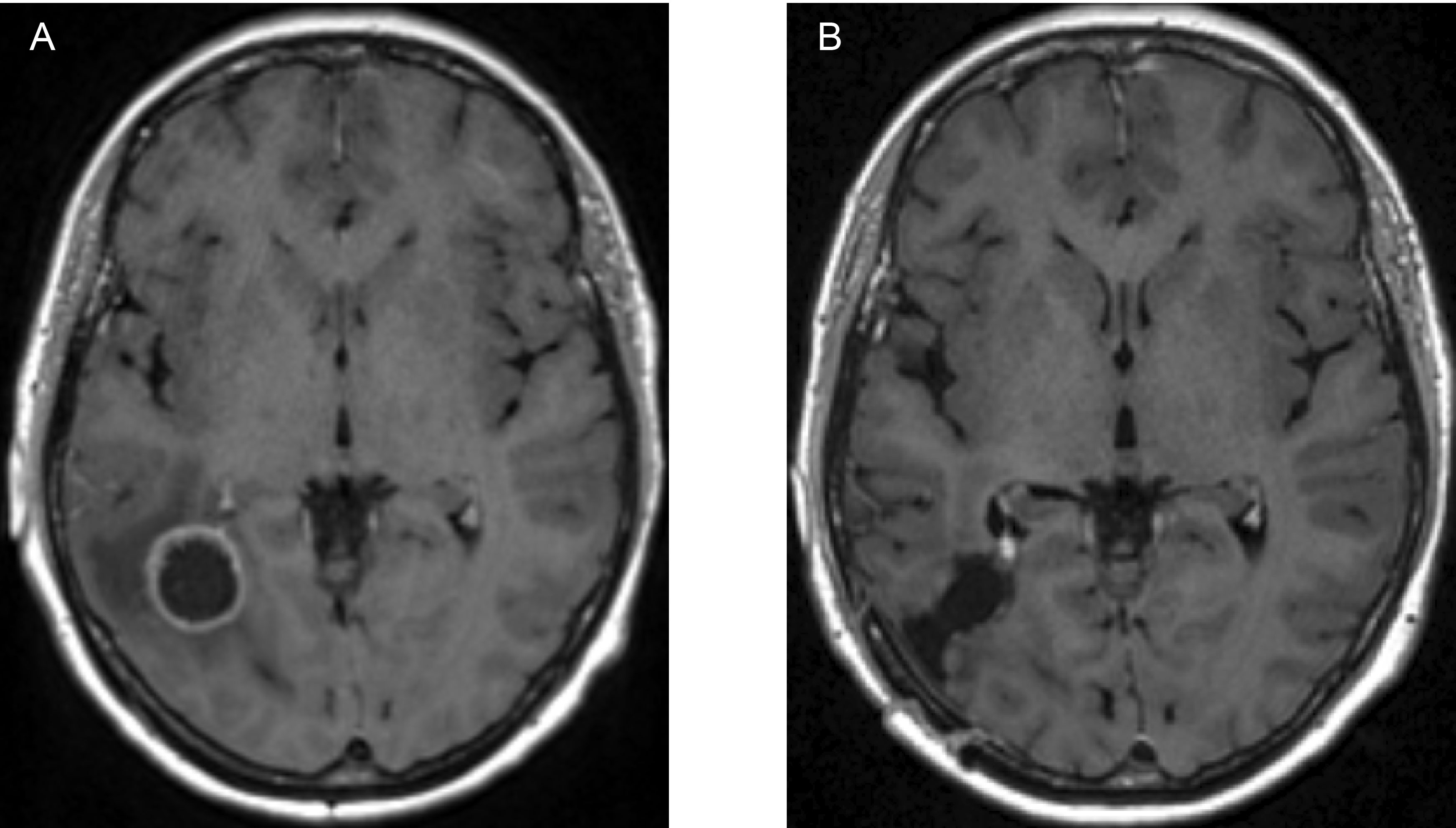} \hfill 
\caption{Image \dg{showing}: \textbf{(A)} axial view of the T1+C pre\textendash operative MRI scan, and \textbf{(B)} axial view of the post\textendash operative MRI scan of the same patient.}
\label{Pre_post_surgery}
\end{figure}

\begin{figure}[ht!]
\centering
\includegraphics[scale=0.5]{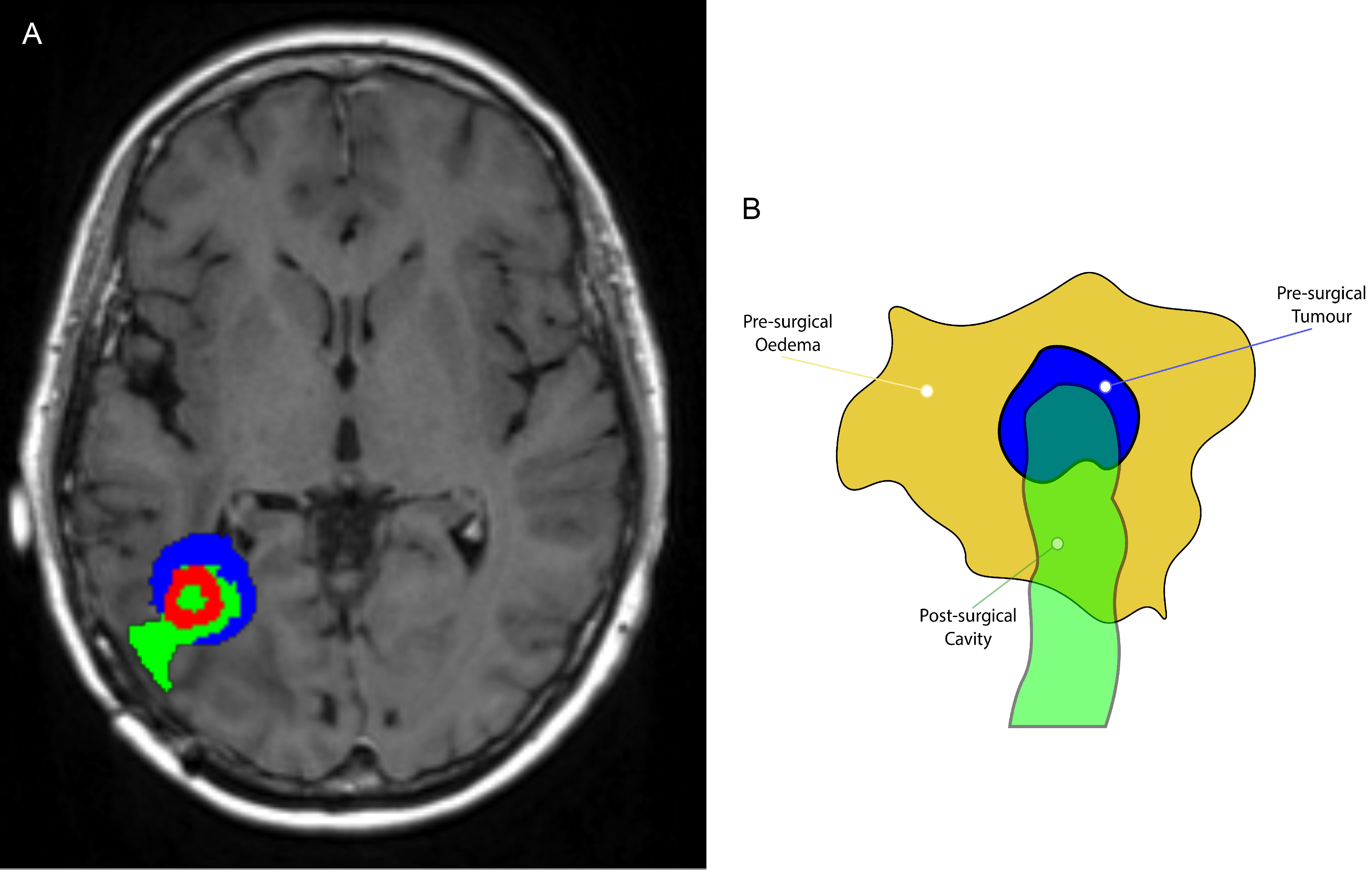} \hfill 
\caption{\textbf{(A)} Superimposition illustrating the spatial alignment of the pre\textendash surgical original tumour (depicted in blue), the post\textendash surgery surgical cavity (highlighted in green), and the recurrent tumour preceding the second surgical procedure (presented in red). \textbf{(B)} \kh{Schematic illustrating the dynamics of the three masks: the oedema mask (in dark yellow), the initial tumour mask (in blue), and the post\textendash surgical mask (in green), with the brighter green region, which is not overlapping with the oedema mask, is designated as zero, \emph{i.e.,} no cancer will be located in this area.}}
\label{Pre&Scan11&Schematic2}
\end{figure}

In earlier stages of our research, we focused solely on the initial oedema volume before surgery and the main tumour size before any operation took place, as shown in Figure \ref{Schematic_1}. However, while this approach was methodologically sound, it falls short when attempting to replicate the evolving changes observed in later MRI scans of the patient. These changes occur as the patient's anatomy undergoes significant transformations due to surgery, as visually depicted in Figure \ref{Pre_post_surgery}.
An illustrative case involves this specific patient who experienced a noticeable reduction in the size of the original tumour site after surgical intervention. This reduction was followed by a recurrence of a smaller tumour, as shown in Figure \ref{Pre&Scan11&Schematic2} A. Consequently, starting our simulations solely based on the initial tumour outline inevitably leads to a tumour size pattern that exceeds our expectations.
%%%%%%%%%%%%%%%%%%%%%%%%%%%%%%%%%%%%%%%%%%%%%%%%%%
%%%%%%%%%%%%%%%%%%%%%%%%%%%%%%%%%%%%%%%%%%%%%%%%%%%
	To address this methodological challenge, we introduced a novel element to our initial conditions: the post\textendash surgical cavity MRI scan. This post\textendash surgical MRI scan provides a clear view of the changes in brain anatomy following surgery, as depicted in Figure \ref{Pre_post_surgery}. By incorporating this post\textendash operative anatomical data into our computational framework, we bridge the crucial gap between the pre\textendash operative and \dg{immediate} post\textendash operative states. This enables a more precise and \dg{anatomically} and physiologically realistic simulation of GBM tumour growth dynamics in the context of surgical interventions.

The mathematical implementation of this innovative volume addition requires a robust framework. We introduce three distinct masks: the oedema mask, the initial tumour mask, and the surgical cavity mask, as illustrated in Figure \ref{Pre&Scan11&Schematic2} B. \amr{We introduce two new constants which indicate the presence or absence of cancer cells within these masks, represented} as $\alpha_{_{it}}$ and $\alpha_{_{sc}}$.  \am{Following the notation described in Section \ref{Reconstruction},} we employ the mollifier distribution within the different masks to articulate this operation as follows:
\am{\begin{itemize}
\item For the initial tumour mask:
$
c_{_{it}}^{d_{_{x}}}(v)\alpha_{_{it}}:= R(n(x),k_{_{R}})^{-1}\psi_{_{1}}\big(\frac{v}{R(n(x),k_{_{R}})}\big), \quad v\in[0,q(n(x),0)], 
$
%\[c_{\text{it}}(\mathbf{x}, t) = \alpha_{\text{it}} \vartheta_{\xi} \ast \mathcal{X}_{(\mathbf{B}(x,R(n(x))))}\]
\item For the surgical cavity mask:
$
c_{_{sc}}^{d_{_{x}}}(v)\alpha_{_{sc}}:= R(n(x),k_{_{R}})^{-1}\psi_{_{1}}\big(\frac{v}{R(n(x),k_{_{R}})}\big), \quad v\in[0,q(n(x),0)]. 
$
\end{itemize}
}
%\[c_{\text{sc}}(\mathbf{x}, t) = \alpha_{\text{sc}} \vartheta_{\xi} \ast \mathcal{X}_{(\mathbf{B}(x,R(n(x))))}\]

%Here, $c_{(\cdot)}(\mathbf{x},t)$ represents the cancer density at spatial point $\mathbf{x}$ and time $t$, $\vartheta_{\xi}$ is the standard mollifier defined previously in \eqref{mollifier}, and $\mathcal{X}_{(\cdot,\cdot)}$ symbolizes the characteristic function for each mask, respectively. 
\am{On the other hand, when using a Gaussian distribution, we have the following equations:
\begin{itemize}
\item For the initial tumour mask:
$
c_{_{it}}^{d_{_{x}}}(v)\alpha_{_{it}}\propto \mathcal{N}_{d_{x}}(0, \tilde{\sigma}(n(x),k_{_{\tilde{\sigma}}})), \qquad v\in[0,q(n(x),0)], 
$
\item For the surgical cavity mask:
$
c_{_{sc}}^{d_{_{x}}}(v)\alpha_{_{sc}}\propto  \mathcal{N}_{d_{x}}(0, \tilde{\sigma}(n(x),k_{_{\tilde{\sigma}}})), \qquad v\in[0,q(n(x),0)]. 
$
\end{itemize}
}
As shown in the schematic diagram in Figure \ref{Pre&Scan11&Schematic2} B , it is clear that the surgical cavity is slightly more elongated than the oedema mask. Therefore, regions where these volumes do not overlap are defined by setting their values to zero. 

In essence, this mathematical framework equips our computational model with the ability to smoothly incorporate the complex interactions between the oedema, initial tumour, and surgical cavity masks. This leads to a more physiologically accurate simulation of GBM tumour growth dynamics, especially in the context of surgical procedures and chemoradiotherapy treatments.

\begin{figure}[ht!]
\centering
\includegraphics[width=\textwidth]{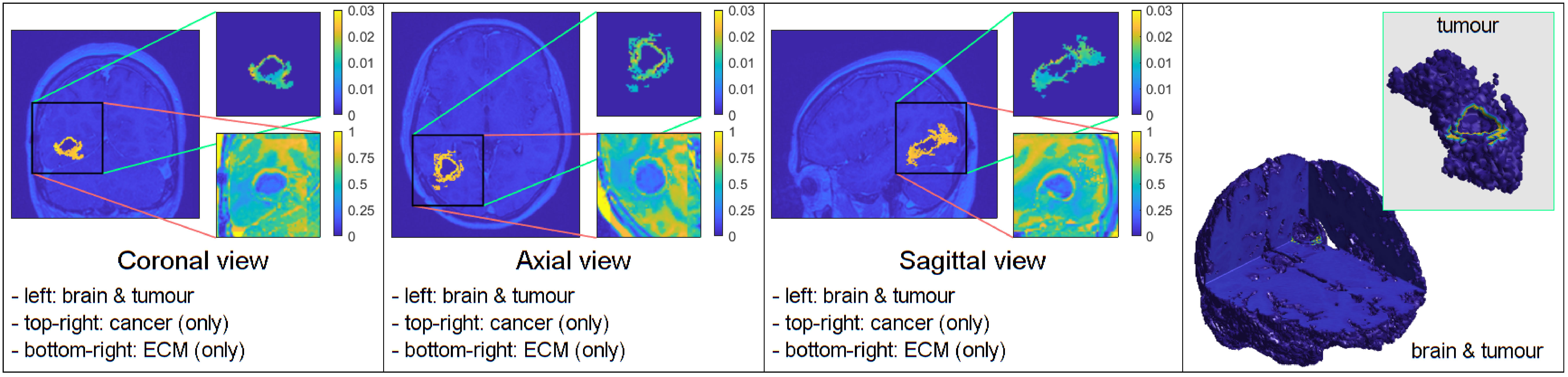} \hfill 
\caption{Simulation with the mollifier distribution using the three introduced masks, with $k_{_{R}}=20$.}
\label{Comparison_1}
\end{figure}

After conducting numerous simulations and carefully adjusting parameters, we obtained simulation results as shown in the illustrative case in Figure \ref{Comparison_1}, where we used the mollifier distribution with $k_{_{R}}=20$ and deactivated the initial tumour mask, \textit{i.e.}, $\alpha_{_{it}}=0$.

\begin{figure}[ht!]
\centering
\includegraphics[scale=0.5]{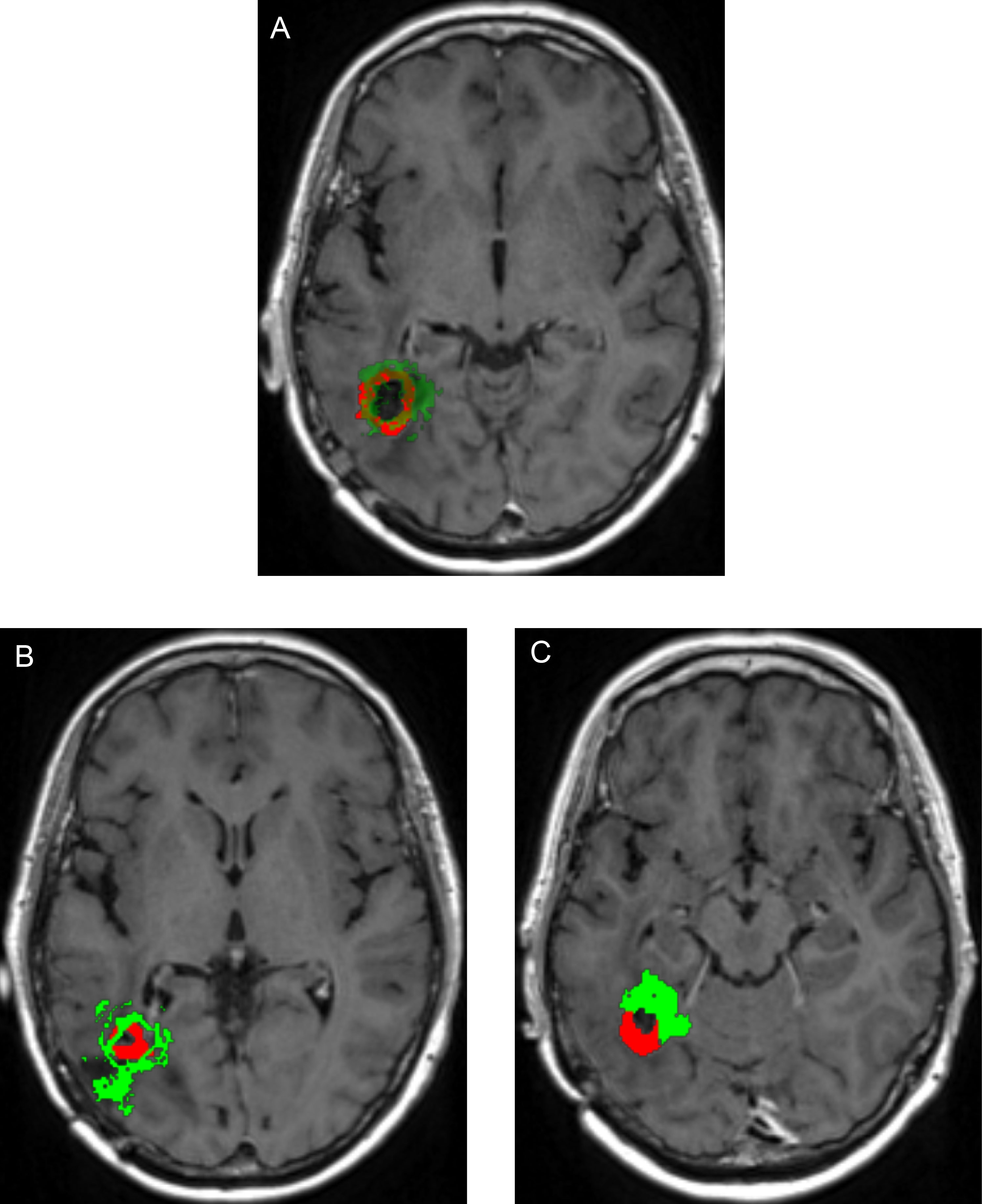} \hfill 
\caption{\textbf{(A)} Overlay depicting the spatial compatibility between our computational simulation (highlighted in green) and the recurred GBM pre\textendash second\textendash surgery MRI scan of the patient (emphasised in red), aligning with the data from the final MRI examination. \textbf{(B-C)} These overlays also involve our computational simulation (in green) and the recurred GBM pre\textendash second\textendash surgery MRI scan (in red). However, in this case, the alignment lacks a high degree of accuracy.}
\label{Comparison_11}
\end{figure}

Notably, the tumour in these results is noticeably smaller and has a more compact spatial distribution. Importantly, this tumour is completely surrounded and confined within the boundaries outlined by the surgical cavity mask, which closely matches the patient's MRI scans, as illustrated in Figure \ref{Comparison_11} A, which represents an overlapping of our simulation (simulation of Figure \ref{Comparison_1} at stage 44, in green) and an MRI scan slice taken 881 days into the treatment, in red. \amr{The simulation of our model closely aligns with real\td world clinical observations for this particular MRI slice. This strong agreement demonstrates the model's effectiveness in predicting relevant outcomes.} \am{While Figure \ref{Comparison_11} A showcases a high degree of accuracy, it is important to acknowledge that not all MRI slices achieve this level of precision, as shown in Figure \ref{Comparison_11} B\td C.} 

This achievement marks a significant milestone in our effort to accurately replicate the complexities of GBM tumour growth in the presence of surgical interventions, treatment administrations, and post\td surgical cancer cell distributions within the oedema. This mathematical modelling contributes to our understanding \dg{of the} clinical management of this \dg{very} challenging medical condition.

\section{Discussion}
\dg{GBM}, an extremely aggressive brain tumour with a low 5\textendash year survival rate of only 7.2\%, poses significant challenges in terms of treatment \citep{burri_2018,Wu_2021}. In the search for better therapies, mathematical modelling has emerged as a valuable approach. Despite established treatments like the Stupp protocol, GBM \kh{almost} \dg{always} recurs, driven by its invasive nature and peritumoural oedema infiltration, in some cases \citep{Heterogeneity_pte, benefit_recurrent, bayesian_meta, pattern_failure}. Mathematical models provide a promising way to understand the complexities of GBM.

Our study \dg{investigates} the connection between the swelling around the tumour (peritumoural oedema) and the distribution of GBM cells within the oedema, whilst using MRI data. Building upon the 3D multiscale moving\textendash boundary framework we introduced earlier, we have incorporated the treatment history of a specific patient from Ninewells Hospital. By simulating how tumours typically grow, our research sets the stage for future experiments using MRI data and treatment histories collected from GBM patients. Ultimately, we aim to develop a mathematical model that incorporates the effects of chemoradiotherapy and investigates the distribution of GBM cells within the oedema with greater accuracy, \amr{whilst also taking into account the anatomical changes of the brain due to surgery.}

In each simulation, we initiate the process by manually segmenting the oedema and pre\textendash surgical tumour masks obtained from the MRI scans of the specific GBM patient. Crucially, we meticulously replicate and \dg{take account of} the exact treatment protocol administered to this patient in our simulations. Furthermore, we investigate two scenarios for how cancer cells are distributed within the oedema: the mollifier and Gaussian distributions. The resulting figures show various outcomes based on different parameter settings. These simulations closely resemble what doctors see in real clinical cases, where recurrent GBM often has the highest concentration of cancer cells at the edge of the surgically removed area. As illustrated in Figures \ref{Results MollifierDist} and \ref{Results GaussianDist}, decreasing $k_{_{R}}$ and $k_{_{\tilde{\sigma}}}$ respectively, corresponds to increased tumour aggressiveness. 

Our experiments, involving a range of parameter combinations and the application of the chemoradiotherapy treatment, have shown that the most controlled and least invasive tumour growth occurs when we start with a maximum cancer cell density of 0.1 and use the mollifier distribution, \amr{with $k_{_{R}}=30$, to arrange the cancer cells within the oedema}, as observed in Figure \ref{BestResults}. 

What is particularly noteworthy is that our simulations closely match MRI scans taken years into the treatment, as shown in Figure \ref{Comparison}, suggesting good agreement between our model and real\textendash world data. This promising finding motivated us to improve our methods to predict GBM growth dynamics by incorporating various pre\textendash and post\textendash operative MRI data along with treatment effects. Initially, we only focused on the initial pre\textendash operative tumour and oedema regions. However, this approach fell short when trying to capture the dynamic changes that occur after surgery, as observed in Figure \ref{Pre_post_surgery}. To address this limitation, we introduced \dg{immediate} post\textendash operative MRI data into our framework, bridging the gap between the pre\textendash operative and post\textendash operative states. This integration involved three masks (the oedema, initial tumour, and a future representation of the surgical cavity) regulated by constants and refined with the \amr{mollifier or Gaussian distributions as necessary}, as shown in Figure \ref{Pre&Scan11&Schematic2} B. This framework allowed for more precise simulations, as evidenced by the reduction in tumour size and spatial distribution, which closely matched the patient's MRI scan taken 881 days after the first surgery, as seen in Figure \ref{Comparison_11} A.

	\am{In conclusion, our model represents a significant advancement in our ability to predict how GBM tumours behave following surgery, treatment administration and the distributions of cancer cells within the oedema. By incorporating pre\textendash operative and post\textendash operative MRI scans and carefully considering patient treatment histories, we have developed a robust framework that accurately replicates the complex dynamics of GBM progression. This achievement not only enhances our understanding of this challenging disease but also opens up significant possibilities in the field of clinical management.}
	
	\amr{However, limitations exist. Even though our simulations closely matched the data, there were some discrepancies, as observed in Figure \ref{Comparison_11} B-C. Further refinement is needed to achieve highly accurate matches. Furthermore, testing on more diverse patients and treatment scenarios is crucial to confirm transferability (ongoing research). Additionally, real\textendash world data from experiments and trials is necessary to refine the parameter values of our model. Addressing these limitations through future research will solidify the model's reliability and effectiveness for real\textendash world applications.}

\am{This research reflects current advancements in GBM research by providing valuable insights into mathematical modelling and its potential to predict this aggressive disease. By translating these insights into improved treatments, we hope this work will lead to a significantly improved outlook for GBM patients.}
%%%%%%%%%%%%%%%%%%%%%%%%%%%%%%%%%%%%%%%%%%%%%%%%%%%%%
\section*{Conflict of Interest Statement}
The authors declare that the research was conducted in the absence of any commercial or financial relationships that could be construed as a potential conflict of interest.
%%%%%%%%%%%%%%%%%%%%%%%%%%%%%%%%%%%%%%%%%%%%%%%%%%%%%%
\section*{Author Contributions}
\amr{ACM, SS, JDS and DT developed and performed the numerical simulations, and  wrote the manuscript. KHI and MO provided all help with clinical data gathering and tumour segmentation. All authors contributed to the article and approved the submitted version.}
%%%%%%%%%%%%%%%%%%%%%%%%%%%%%%%%%%%%%%%%%%%%%%%%%%%%%%
\section*{Ethics statement}
Ethical approval was obtained from the local Caldicott Guardian, Integrated Research Application System (IRAS)(project ID: 309957), Tayside Research and Development Committee (project ID: 2022NH01) and Research Ethics Committee (REC) (Ref: 22/NS/0021).
%%%%%%%%%%%%%%%%%%%%%%%%%%%%%%%%%%%%%%%%%%%%%%%%%%%%%%%
\section*{Funding}
\amr{ACM, DT, JDS and KHI would like to acknowledge the generous funding received from the Ninewells Cancer Campaign (NCC) - Fraser Fellowships Doctoral Training Programme (DTP) in Precision Cancer Medicine, which fully funds this project.}
%%%%%%%%%%%%%%%%%%%%%%%%%%%%%%%%%%%%%%%%%%%%%%%%%%%%%%%%
\section*{Acknowledgements}
\amr{We would like to acknowledge the generous help provided by Dr. Jennifer MacFarlane in gathering the GBM data.}
%%%%%%%%%%%%%%%%%%%%%%%%%%%%%%%%%%%%%%%%%%%%%%%%%%%%%%%%
\section*{Data Availability Statement}
\amr{Permission for access to the data is available via request to the NHS Tayside Caldicott Guardian, and NHS Tayside Ethics and R\&D Department.}  
%%%%%%%%%%%%%%%%%%APPENDIX Sections
\appendix
\section*{APPENDIX}
\section{Outside Tumour Boundary}
\label{outTumourBry}
%%%%%%%%%%%%%%%%%%%%%%%%%%%%%%%%%%%%%%%%%%%%%%%%%%%%%%%%
Following the definition of $\partial \Omega_{o}(t)$ by \citet{Szabolcs_2022_Nutrients}, let $x \in \partial \Omega (t)$. Then, $x\in \partial \Omega_{o}(t)$ if and only if there exists $\phi_x:[0,\infty)\rightarrow \R^d$ such that the following properties hold true simultaneously:
\begin{equation*}
\begin{array}{lll}
&1)& \phi_x(0)=x,\\ [0.2cm]
&2)& \phi_x(s)\neq x, \forall s \in (0,\infty),\\ [0.2cm]
&3)& Im\phi_x \backslash \{x\} \subset \complement \Omega(t),\\ [0.2cm]
&4)& \lim\limits_{s \to \infty} dist(\phi(s),\partial \Omega(t))=\infty,
\end{array}
\end{equation*}
where $\forall s \in (0,\infty),$ we have $ dist(\phi(s),\partial \Omega(t)):= \inf\limits_{x\in \partial \Omega(t)} \Vert\phi(s)-z\Vert_{_{2}}$ and represents the Euclidean distance from $\phi(s)$ to $\partial \Omega(t)$.
\begin{figure}[h!]
\centering
\includegraphics[scale=0.5]{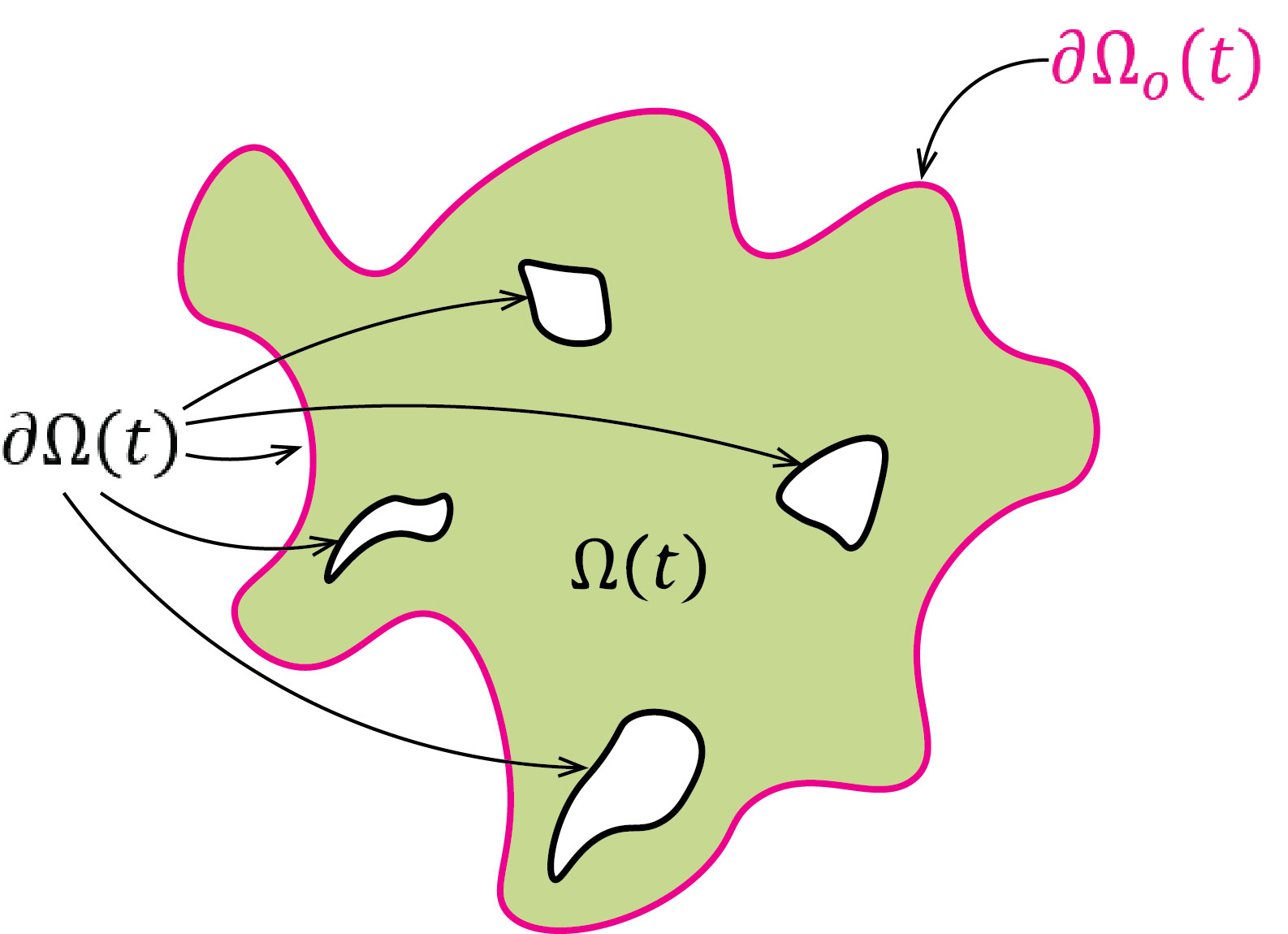} \hfill 
\caption{Schematic showing $\partial \Omega_{o}(t)$, the outer boundary highlighted with the dashed line.}
\label{Outside_Boundary}
\end{figure}
%%%%%%%%%%%%%%%%%%%%%%%%%%%%%%%%%%%%%%%%%%%%%%%%%
\section{Table of Parameters}
Here we include the table with the baseline values for the parameters considered for our model at macro\textendash scale. Furthermore, any other choices in the parameter values (\emph{i.e.,} beyond the ones included in this table) is explained as appropriate in the text.
%%%%%%%%%%%%%%%%%%%%%%%%%%%%%%%%%%%%%%%%%%%%%%%
\begin{table}[h!]
\centering
\begin{tabular}{|l|c|c|}
 \hline
 Parameters & Values & References \\ 
 \hline 
 $D_{c}$ & $10^{-4}$ & \citep{Szabolcs_2022_Nutrients} \\
 \hline
 $D_G$ & $0.25$ & \citep{suveges_2021_mathematical} \\  
 \hline 
 $D_{\sigma}$ & $1$ & \citep{Szabolcs_2022_Nutrients} \\
 \hline
 $D_{m}$ & $2.5 \times 10^{-3}$ & \citep{Szabolcs_2022_Nutrients} \\
 \hline
 $\mathcal{K}_{FA}$ & $100$ & \citep{suveges_2021_mathematical} \\
 \hline
 $\mathbf{S}_{\max}$ & $0.5$ & \citep{Szabolcs_2022_Nutrients} \\
 \hline
  $\mathbf{S}_{\min}$ & $0.01$ & \citep{Szabolcs_2022_Nutrients} \\  
 \hline 
  $\mathbf{S}_{cl}$ & $0.01$ & \citep{Szabolcs_2022_Nutrients} \\ 
 \hline
  $\mathbf{S}_{cF}$ & $0.3$ & \citep{Szabolcs_2022_Nutrients} \\ 
 \hline
 $\mu$ & $0.25$ & \citep{Szabolcs_2022_Nutrients} \\
 \hline
  $d$& $0.015$ & \citep{Szabolcs_2022_Nutrients} \\  
 \hline 
 $d_{\sigma}$ & $80$ & \citep{Szabolcs_2022_Nutrients} \\
 \hline
 $\beta_l$ & $3$ & \citep{Szabolcs_2022_Nutrients} \\
 \hline
 $\beta_F$ & $1.5$ & \citep{Szabolcs_2022_Nutrients} \\
 \hline
   $\sigma_{nor}$ & $0.4$ & \citep{Szabolcs_2022_Nutrients} \\  
 \hline 
 $\sigma_p$ & $0.35$ & \citep{Szabolcs_2022_Nutrients} \\
 \hline 
 $\sigma_n$ & $0.2$ & \citep{Szabolcs_2022_Nutrients} \\
 \hline
 $\Psi_{p,\max}$ & $1$ & \citep{Szabolcs_2022_Nutrients} \\
 \hline
 $\Psi_{d,\max}$ & $5$ & \citep{Szabolcs_2022_Nutrients} \\
 \hline
 $b$ & $1.1$ & \citep{Irina_2021} \\
 \hline
 $s$ & $2$ & \citep{Irina_2021} \\
 \hline
 $\beta$ & $0.5$ & \citep{Irina_2021} \\
 \hline
 $W$ & $130$ & Clinical estimate \\
 \hline
 $\alpha$ & $10.2$ & \citep{alpha_beta} \\
 \hline
 $\zeta$ & $8$ & \citep{alpha_beta} \\
 \hline
 $\beta_{FChemo}$ & $0.5$ & Estimated \\
 \hline
 $\beta_{FRadio}$ & $0.5$ & Estimated \\
 \hline
 $\beta_{lChemo}$ & $0.5$ & Estimated \\
 \hline
 $\beta_{lRadio}$ & $0.5$ & Estimated \\
 \hline
 $R$ & $0.15$ & \citep{Szabolcs_2022_Nutrients} \\
 \hline
 $r$ & $0.0016$ & \citep{Szabolcs_2022_Nutrients} \\
 \hline
 $f_{\max}$ & $0.636$ & \citep{Szabolcs_2022_Nutrients} \\
 \hline 
  $N_{_{radio}}$ & $30$ & Clinical estimate\\
  \hline
   $N_{_{chemo}}$ & $132$ & Clinical estimate\\
  \hline
  $\bar D(j_{_{m}}), \, \forall m=1\dots N_{_{radio}}$ & $2Gy$ & Clinical estimate\\
  \hline
\end{tabular}\\[0.2cm]
\caption{The baseline parameter values used for the numerical simulations.}
\label{Table_1}
\end{table}
%%%%%%%%%%%%%%%%%%%%%%%%%%%%%%%%%%%%%%%%%%%%%%%%%%%%%%%%%%%%%%%%%%%%%%%%%%%%
\section{The Standard Mollifier and the per-Day Radio and Chemo Scheduling}
\label{mollifierAppendix}
The form of the standard symmetric mollifier on $\psi_{_{n}}:\R^{n}\to \R_{+}$, $n\in\{1,3\}$, used in this manuscript is given by:
\begin{equation}\label{mollifier}
\begin{array}{rl}
 \psi_{_{n}}(x)&=
 \left\{
 \begin{array}{ll}
 \exp\Bigg(\frac{-1}{1-\Vert \mathbf{x} \Vert^{^{2}}_{_{2}}}\Bigg)&, \quad x \in \mathbf{B}(0,1)  \\[0.2cm]
 0&, \quad x \notin \mathbf{B}(0,1), 
 \end{array}
 \right.\\[0.2cm]
\end{array} 
\end{equation}
Finally, the overlapping effect for both chemo- and radio- therapy delivery, is described by
\bequ\label{schedulingFunc}
\begin{array}{ll}
\psi^{chemo}_{{i_{_{k}}}}(t)=\psi^{scheduling}(i_{_{k}},t),&\quad \forall k\in\{1\dots N_{_{chemo}}\},\\[0.2cm]
\psi^{radio}_{{j_{_{m}}}}(t)=\psi^{scheduling}(j_{_{m}},t),&\quad \forall m\in\{1\dots N_{_{radio}}\},
\end{array}
\eequ
with
\begin{align*}
\psi^{scheduling}(p,t)&:=
\begin{cases} 
e^{\Bigl(\frac{1}{d^2}-\frac{1}{d^2-(t-T_{{p}})^2}\Bigr)} & \mbox{if } t\in (T_{{p}},T_{{p}}+d), \\ 
e^{\Bigl(\frac{1}{l^2}-\frac{1}{l^2-(t-T_{{p}})^2}\Bigr)} & \mbox{if } t\in (T_{{p}}-l,T_{{p}}),\\ 0 & \mbox{if } t\in (-\infty,T_{{p}}-l) \cup (T_{{p}}+d, +\infty),
\end{cases}
\end{align*} 
%%%%%%%%%%%%%%%%%%%%% BIBLIOGRAPHY
\bibliographystyle{Frontiers-Harvard} 
\bibliography{References}

\begin{thebibliography}{43}
\providecommand{\natexlab}[1]{#1}
\expandafter\ifx\csname urlstyle\endcsname\relax
  \providecommand{\doi}[1]{doi:\discretionary{}{}{}#1}\else
  \providecommand{\doi}{doi:\discretionary{}{}{}\begingroup
  \urlstyle{rm}\Url}\fi
\providecommand{\selectlanguage}[1]{\relax}
\providecommand{\bibAnnoteFile}[1]{%
  \IfFileExists{#1}{\begin{quotation}\noindent\textsc{Key:} #1\\
  \textsc{Annotation:}\ \input{#1}\end{quotation}}{}}
\providecommand{\bibAnnote}[2]{%
  \begin{quotation}\noindent\textsc{Key:} #1\\
  \textsc{Annotation:}\ #2\end{quotation}}

\bibitem[{Alzahrani et~al.(2019)Alzahrani, Eftimie, and
  Trucu}]{alzahrani_2019_multiscale}
Alzahrani, T., Eftimie, R., and Trucu, D. (2019).
\newblock Multiscale modelling of cancer response to oncolytic viral therapy.
\newblock \emph{Mathematical Biosciences} 310, 76--95.
\newblock \doi{10.1016/j.mbs.2018.12.018}
\bibAnnoteFile{alzahrani_2019_multiscale}

\bibitem[{Bashkirtseva et~al.(2021)Bashkirtseva, Ryashko, L{\'{o}}pez, Seoane,
  and Sanju{\'{a}}n}]{Irina_2021}
Bashkirtseva, I., Ryashko, L., L{\'{o}}pez, {\'{A}}.~G., Seoane, J.~M., and
  Sanju{\'{a}}n, M. A.~F. (2021).
\newblock The effect of time ordering and concurrency in a mathematical model
  of chemoradiotherapy 96, 105693.
\newblock \doi{10.1016/j.cnsns.2021.105693}
\bibAnnoteFile{Irina_2021}

\bibitem[{Brooks et~al.(2021)Brooks, Clements, Burden, Kocher, Richards, Devesa
  et~al.}]{Brooks_2021}
Brooks, L.~J., Clements, M.~P., Burden, J.~J., Kocher, D., Richards, L.,
  Devesa, S.~C., et~al. (2021).
\newblock The white matter is a pro-differentiative niche for glioblastoma 12.
\newblock \doi{10.1038/s41467-021-22225-w}
\bibAnnoteFile{Brooks_2021}

\bibitem[{Burri et~al.(2018)Burri, Gondi, Brown, and Mehta}]{burri_2018}
Burri, S.~H., Gondi, V., Brown, P.~D., and Mehta, M.~P. (2018).
\newblock The evolving role of tumor treating fields in managing glioblastoma:
  Guide for oncologists.
\newblock \emph{American Journal of Clinical Oncology} 41, 191--196.
\newblock \doi{10.1097/COC.0000000000000395}
\bibAnnoteFile{burri_2018}

\bibitem[{Chen et~al.(2021)Chen, Wang, Zhao, and et~al.}]{bayesian_meta}
Chen, W., Wang, Y., Zhao, B., and et~al. (2021).
\newblock Optimal therapies for recurrent glioblastoma: A bayesian network
  meta-analysis.
\newblock \emph{Frontiers in Oncology} 11.
\newblock \doi{10.3389/fonc.2021.641878}
\bibAnnoteFile{bayesian_meta}

\bibitem[{Chicoine and Silbergeld(1995)}]{Chicoine_1995}
Chicoine, M.~R. and Silbergeld, D.~L. (1995).
\newblock Assessment of brain tumor cell motility in vivo and in vitro 82,
  615--622.
\newblock \doi{10.3171/jns.1995.82.4.0615}
\bibAnnoteFile{Chicoine_1995}

\bibitem[{Engwer et~al.(2014)Engwer, Hillen, Knappitsch, and
  Surulescu}]{engwer_2014_glioma}
Engwer, C., Hillen, T., Knappitsch, M., and Surulescu, C. (2014).
\newblock Glioma follow white matter tracts: a multiscale dti-based model.
\newblock \emph{Journal of Mathematical Biology} 71, 551--582.
\newblock \doi{10.1007/s00285-014-0822-7}
\bibAnnoteFile{engwer_2014_glioma}

\bibitem[{Hanahan(2022)}]{Hallmarks_Cancer_2022}
Hanahan, D. (2022).
\newblock {Hallmarks of Cancer: New Dimensions}.
\newblock \emph{Cancer Discovery} 12, 31--46.
\newblock \doi{10.1158/2159-8290.CD-21-1059}
\bibAnnoteFile{Hallmarks_Cancer_2022}

\bibitem[{Hatzikirou et~al.(2005)Hatzikirou, Deutsch, Schaller, Simon, and
  et~al.}]{hatzikirou_2005_mathematical}
Hatzikirou, H., Deutsch, A., Schaller, C., Simon, and et~al. (2005).
\newblock Mathematical modelling of glioblastoma tumour development: a review.
\newblock \emph{Mathematical Models and Methods in Applied Sciences} 15,
  1779--1794.
\newblock \doi{10.1142/s0218202505000960}
\bibAnnoteFile{hatzikirou_2005_mathematical}

\bibitem[{Hillen et~al.(2017)Hillen, J.~Painter, C.~Swan, and
  D.~Murtha}]{ref65}
Hillen, T., J.~Painter, K., C.~Swan, A., and D.~Murtha, A. (2017).
\newblock Moments of von mises and fisher distributions and applications 14,
  673--694.
\newblock \doi{10.3934/mbe.2017038}
\bibAnnoteFile{ref65}

\bibitem[{{IXI Dataset}(2024)}]{ref53}
[Dataset] {IXI Dataset} (2024).
\newblock {IXI} brain imaging dataset.
\newblock IXI Repository
\bibAnnoteFile{ref53}

\bibitem[{Jiang and Shu(1996)}]{Jiang_1996}
Jiang, G.-S. and Shu, C.-W. (1996).
\newblock Efficient implementation of weighted eno schemes.
\newblock \emph{Journal of Computational Physics} 126, 202--228.
\newblock \doi{10.1006/jcph.1996.0130}
\bibAnnoteFile{Jiang_1996}

\bibitem[{Kim and Kwon(2005)}]{Kim_2005}
Kim, D. and Kwon, J.~H. (2005).
\newblock A high-order accurate hybrid scheme using a central flux scheme and a
  {WENO} scheme for compressible flowfield analysis.
\newblock \emph{Journal of Computational Physics} 210, 554--583.
\newblock \doi{10.1016/j.jcp.2005.04.023}
\bibAnnoteFile{Kim_2005}

\bibitem[{L{\^e} et~al.(2017)L{\^e}, Delingette, Kalpathy-Cramer, Gerstner,
  Batchelor, Unkelbach et~al.}]{Le_2017}
L{\^e}, M., Delingette, H., Kalpathy-Cramer, J., Gerstner, E.~R., Batchelor,
  T., Unkelbach, J., et~al. (2017).
\newblock Personalized radiotherapy planning based on a computational tumor
  growth model.
\newblock \emph{IEEE Transactions on Medical Imaging} 36, 815--825.
\newblock \doi{10.1109/TMI.2016.2626443}
\bibAnnoteFile{Le_2017}

\bibitem[{Lem{\'e}e(2015)}]{Heterogeneity_pte}
Lem{\'e}e, J.-M. e.~a. (2015).
\newblock Intratumoral heterogeneity in glioblastoma: don't forget the
  peritumoral brain zone.
\newblock \emph{Neuro-oncology} 17, 1322--1332.
\newblock \doi{10.1093/neuonc/nov119}
\bibAnnoteFile{Heterogeneity_pte}

\bibitem[{Lipkova et~al.(2019)Lipkova, Angelikopoulos, Wu, and
  et~al.}]{Radio_Bayesian_GBM}
Lipkova, J., Angelikopoulos, P., Wu, S., and et~al. (2019).
\newblock Personalized radiotherapy design for glioblastoma: Integrating
  mathematical tumor models, multimodal scans, and bayesian inference.
\newblock \emph{IEEE Transactions on Medical Imaging} 38, 1875--1884.
\newblock \doi{10.1109/TMI.2019.2902044}
\bibAnnoteFile{Radio_Bayesian_GBM}

\bibitem[{Liu et~al.(1994)Liu, Osher, and Chan}]{Liu_1994}
Liu, X.-D., Osher, S., and Chan, T. (1994).
\newblock Weighted essentially non-oscillatory schemes.
\newblock \emph{Journal of Computational Physics} 115, 200--212.
\newblock \doi{10.1006/jcph.1994.1187}
\bibAnnoteFile{Liu_1994}

\bibitem[{Malinzi et~al.(2017)Malinzi, Eladdadi, and
  Sibanda}]{malinzi_2017_modelling}
Malinzi, J., Eladdadi, A., and Sibanda, P. (2017).
\newblock Modelling the spatiotemporal dynamics of chemovirotherapy cancer
  treatment.
\newblock \emph{Journal of Biological Dynamics} 11, 244--274.
\newblock \doi{10.1080/17513758.2017.1328079}
\bibAnnoteFile{malinzi_2017_modelling}

\bibitem[{Mardia and Jupp(1999)}]{ref66}
Mardia, K.~V. and Jupp, P.~E. (1999).
\newblock \emph{Directional Statistics} (Wiley).
\newblock \doi{10.1002/9780470316979}
\bibAnnoteFile{ref66}

\bibitem[{Michor and Beal(2015)}]{michor_2015}
Michor, F. and Beal, K. (2015).
\newblock Improving cancer treatment via mathematical modeling: Surmounting the
  challenges is worth the effort.
\newblock \emph{Cell} 163(5), 1059--1063.
\newblock \doi{10.1016/j.cell.2015.11.002}
\bibAnnoteFile{michor_2015}

\bibitem[{Mizuhata et~al.(2023)Mizuhata, Takamatsu, Shibata, and
  et~al.}]{pattern_failure}
Mizuhata, M., Takamatsu, S., Shibata, S., and et~al. (2023).
\newblock Patterns of failure in glioblastoma multiforme following standard (60
  gy) or short course (40 gy) radiation and concurrent temozolomide.
\newblock \emph{Japanese journal of radiology} 41, 660--668.
\newblock \doi{10.1007/s11604-023-01386-2}
\bibAnnoteFile{pattern_failure}

\bibitem[{Niyazi et~al.(2023)Niyazi, Andratschke, Bendszus, Chalmers, Erridge,
  Galldiks et~al.}]{Niyazi_2023}
Niyazi, M., Andratschke, N., Bendszus, M., Chalmers, A.~J., Erridge, S.~C.,
  Galldiks, N., et~al. (2023).
\newblock Estro-eano guideline on target delineation and radiotherapy details
  for glioblastoma 184, 109663.
\newblock \doi{10.1016/j.radonc.2023.109663}
\bibAnnoteFile{Niyazi_2023}

\bibitem[{Oh et~al.(2011)Oh, Sahgal, Sanghera, and
  et~al.}]{recurrence2011patterns}
Oh, J., Sahgal, A., Sanghera, P., and et~al. (2011).
\newblock Glioblastoma: patterns of recurrence and efficacy of salvage
  treatments.
\newblock \emph{The Canadian journal of neurological sciences.} 38, 621--625.
\newblock \doi{10.1017/s0317167100012166}
\bibAnnoteFile{recurrence2011patterns}

\bibitem[{Painter and Hillen(2013)}]{ref37}
Painter, K.~J. and Hillen, T. (2013).
\newblock Mathematical modelling of glioma growth: The use of diffusion tensor
  imaging (dti) data to predict the anisotropic pathways of cancer invasion
  323, 25--39.
\newblock \doi{10.1016/j.jtbi.2013.01.014}
\bibAnnoteFile{ref37}

\bibitem[{Petrecca et~al.(2013)Petrecca, Guiot, Panet-Raymond, and
  Souhami}]{petrecca2013failure}
Petrecca, K., Guiot, M.-C., Panet-Raymond, V., and Souhami, L. (2013).
\newblock Failure pattern following complete resection plus radiotherapy and
  temozolomide is at the resection margin in patients with glioblastoma.
\newblock \emph{Journal of neuro-oncology} \doi{10.1007/s11060-012-0983-4}
\bibAnnoteFile{petrecca2013failure}

\bibitem[{Plaszczynski et~al.(2023)Plaszczynski, Grammaticos, Pallud, Campagne,
  and Badoual}]{Plaszczynski_2023}
Plaszczynski, S., Grammaticos, B., Pallud, J., Campagne, J.-E., and Badoual, M.
  (2023).
\newblock Predicting regrowth of low-grade gliomas after radiotherapy.
\newblock \emph{PLOS Computational Biology} 19, 1--16.
\newblock \doi{10.1371/journal.pcbi.1011002}
\bibAnnoteFile{Plaszczynski_2023}

\bibitem[{Qin et~al.(2021)Qin, Liu, Akter, Qin, Xie, Li et~al.}]{Qin_2021}
Qin, X., Liu, R., Akter, F., Qin, F., Xie, Q., Li, Y., et~al. (2021).
\newblock Peri-tumoral brain edema associated with glioblastoma correlates with
  tumor recurrence.
\newblock \emph{Journal of Cancer} 12(7), 2073--2082.
\newblock \doi{10.7150/jca.53198}
\bibAnnoteFile{Qin_2021}

\bibitem[{Ringel et~al.(2016)Ringel, Pape, Sabel, and
  et~al.}]{benefit_recurrent}
Ringel, F., Pape, H., Sabel, M., and et~al. (2016).
\newblock Clinical benefit from resection of recurrent glioblastomas: results
  of a multicenter study including 503 patients with recurrent glioblastomas
  undergoing surgical resection.
\newblock \emph{Neuro-Oncology} 18, 96--104.
\newblock \doi{10.1093/neuonc/nov145}
\bibAnnoteFile{benefit_recurrent}

\bibitem[{Rockne et~al.(2010)Rockne, Rockhill, Mrugala, and
  et~al.}]{Rockne_2010}
Rockne, R., Rockhill, J.~K., Mrugala, M., and et~al. (2010).
\newblock Predicting the efficacy of radiotherapy in individual glioblastoma
  patients in vivo: a mathematical modeling approach.
\newblock \emph{Phys. Med. Biol.} 55, 3271--3285.
\newblock \doi{10.1088/0031-9155/55/12/001}
\bibAnnoteFile{Rockne_2010}

\bibitem[{Shuttleworth and Trucu(2019)}]{shuttleworth_2019_multiscale}
Shuttleworth, R. and Trucu, D. (2019).
\newblock Multiscale modelling of fibres dynamics and cell adhesion within
  moving boundary cancer invasion.
\newblock \emph{Bulletin of Mathematical Biology} 81, 2176--2219.
\newblock \doi{10.1007/s11538-019-00598-w}
\bibAnnoteFile{shuttleworth_2019_multiscale}

\bibitem[{Shuttleworth and Trucu(2020)}]{shuttleworth_2020}
Shuttleworth, R. and Trucu, D. (2020).
\newblock Cell-scale degradation of peritumoural extracellular matrix fibre
  network and its role within tissue-scale cancer invasion.
\newblock \emph{Bull Math Biol} 82.
\newblock \doi{10.1007/s11538-020-00732-z}
\bibAnnoteFile{shuttleworth_2020}

\bibitem[{Silbergeld and Chicoine(1997)}]{Silbergeld_1997}
Silbergeld, D.~L. and Chicoine, M.~R. (1997).
\newblock Isolation and characterization of human malignant glioma cells from
  histologically normal brain 86, 525--531.
\newblock \doi{10.3171/jns.1997.86.3.0525}
\bibAnnoteFile{Silbergeld_1997}

\bibitem[{Stupp et~al.(2005)Stupp, Mason, van~den Bent, Weller, Fisher,
  Taphoorn et~al.}]{Stupp_2005}
Stupp, R., Mason, W.~P., van~den Bent, M.~J., Weller, M., Fisher, B., Taphoorn,
  M. J.~B., et~al. (2005).
\newblock Radiotherapy plus concomitant and adjuvant temozolomide for
  glioblastoma 352, 987--996.
\newblock \doi{10.1056/NEJMoa043330}
\bibAnnoteFile{Stupp_2005}

\bibitem[{Suveges et~al.(2022)Suveges, Eftimie, and
  Trucu}]{Szabolcs_2022_Nutrients}
Suveges, S., Eftimie, R., and Trucu, D. (2022).
\newblock Re-polarisation of macrophages within collective tumour cell
  migration: A multiscale moving boundary approach.
\newblock \emph{Frontiers in Applied Mathematics and Statistics} 7.
\newblock \doi{10.3389/fams.2021.799650}
\bibAnnoteFile{Szabolcs_2022_Nutrients}

\bibitem[{Suveges et~al.(2021)Suveges, Hossain-Ibrahim, Steele, Eftimie, and
  Trucu}]{suveges_2021_mathematical}
Suveges, S., Hossain-Ibrahim, K., Steele, J.~D., Eftimie, R., and Trucu, D.
  (2021).
\newblock Mathematical modelling of glioblastomas invasion within the brain: A
  3d multi-scale moving-boundary approach.
\newblock \emph{Mathematics} 9, 2214.
\newblock \doi{10.3390/math9182214}
\bibAnnoteFile{suveges_2021_mathematical}

\bibitem[{Swanson(2008)}]{swanson_2008}
Swanson, K. e.~a. (2008).
\newblock A mathematical modelling tool for predicting survival of individual
  patients following resection of glioblastoma: a proof of principle.
\newblock \emph{Br J Cancer} 98, 113--119.
\newblock \doi{10.1038/sj.bjc.6604125}
\bibAnnoteFile{swanson_2008}

\bibitem[{Swanson et~al.(2000)Swanson, Alvord, and Murray}]{Swanson_2000}
Swanson, K.~R., Alvord, E.~C., and Murray, J.~D. (2000).
\newblock A quantitative model for differential motility of gliomas in grey and
  white matter 33, 317--329.
\newblock \doi{10.1046/j.1365-2184.2000.00177.x}
\bibAnnoteFile{Swanson_2000}

\bibitem[{Trucu et~al.(2013)Trucu, Lin, Chaplain, and
  Wang}]{trucu_2013_multiscale}
Trucu, D., Lin, P., Chaplain, M. A.~J., and Wang, Y. (2013).
\newblock A multiscale moving boundary model arising in cancer invasion.
\newblock \emph{Multiscale Modeling \& Simulation} 11, 309--335.
\newblock \doi{10.1137/110839011}
\bibAnnoteFile{trucu_2013_multiscale}

\bibitem[{van Leeuwen et~al.(2018)van Leeuwen, Oei, and Crezee}]{alpha_beta}
van Leeuwen, C.~M., Oei, A.~L., and Crezee, J. e.~a. (2018).
\newblock The alfa and beta of tumours: a review of parameters of the
  linear-quadratic model, derived from clinical radiotherapy studies.
\newblock \emph{Radiation Oncology} 13, 96.
\newblock \doi{10.1186/s13014-018-1040-z}
\bibAnnoteFile{alpha_beta}

\bibitem[{Wu et~al.(2021)Wu, Klockow, Zhang, Lafortune, Chang, Jin
  et~al.}]{Wu_2021}
Wu, W., Klockow, J.~L., Zhang, M., Lafortune, F., Chang, E., Jin, L., et~al.
  (2021).
\newblock Glioblastoma multiforme (gbm): An overview of current therapies and
  mechanisms of resistance.
\newblock \emph{Pharmacological Research} 171, 105780.
\newblock \doi{10.1016/j.phrs.2021.105780}
\bibAnnoteFile{Wu_2021}

\bibitem[{Yalamarty et~al.(2023)Yalamarty, Filipczak, Li, Subhan, Parveen,
  Ataide et~al.}]{Yalamarty_2023}
Yalamarty, S. S.~K., Filipczak, N., Li, X., Subhan, M.~A., Parveen, F., Ataide,
  J.~A., et~al. (2023).
\newblock Mechanisms of resistance and current treatment options for
  glioblastoma multiforme (gbm) 15, 2116.
\newblock \doi{10.3390/cancers15072116}
\bibAnnoteFile{Yalamarty_2023}

\bibitem[{Yin and Guchelaar(2019)}]{Review_Solid_Tumours}
Yin, M. D. J. A. R. v. H. J. G. C. S. J.~J., A. and Guchelaar, H.-J. (2019).
\newblock A review of mathematical models for tumor dynamics and treatment
  resistance evolution of solid tumors.
\newblock \emph{CPT Pharmacometrics Syst. Pharmacol.} 8, 720--737.
\newblock \doi{10.1002/psp4.12450}
\bibAnnoteFile{Review_Solid_Tumours}

\bibitem[{Zhang and Shu(2006)}]{Zhang_2006}
Zhang, S. and Shu, C.-W. (2006).
\newblock A new smoothness indicator for the {WENO} schemes and its effect on
  the convergence to steady state solutions.
\newblock \emph{Journal of Scientific Computing} 31, 273--305.
\newblock \doi{10.1007/s10915-006-9111-y}
\bibAnnoteFile{Zhang_2006}

\end{thebibliography}
\end{document}